\documentclass[sigconf,balance=false]{acmart}
\usepackage{popets}
\usepackage{float}
\usepackage{booktabs}
\usepackage{graphicx}
\usepackage{multirow}

\usepackage{xspace}
\usepackage{amsmath}
\usepackage{arabtex}

\usepackage{utf8}
\usepackage[T1]{fontenc}
\usepackage[utf8]{inputenc}
\usepackage{xspace} % Smart spacing
\usepackage{balance}
\usepackage{enumitem}
\usepackage{mathptmx}
\usepackage{graphicx}
\usepackage{array}
\usepackage{xparse}
\usepackage{amsmath}
\usepackage{anyfontsize}
\usepackage{color}
\usepackage{rotating}
\usepackage{adjustbox}
\usepackage{colortbl}
\usepackage{microtype}
\usepackage{etoolbox}
\usepackage{tikz}
\usepackage[english]{babel}
\usepackage{multirow}

\usepackage{tabularx}

\makeatletter
\patchcmd{\@makecaption}
  {\scshape}
  {}
  {}
  {}
\makeatother
	
\definecolor{lightcyan}{rgb}{0.88,1,1}
\definecolor{cadetgrey}{rgb}{0.57, 0.64, 0.69}
\definecolor{airforceblue}{rgb}{0.36, 0.64, 0.66}

\newrobustcmd*{\mycircle}[1]{\tikz{\filldraw[draw=#1,fill=#1] (0,0) circle [radius=0.1cm];}}

\newrobustcmd*{\mytriangle}[1]{\tikz{\filldraw[draw=#1,fill=#1] (0,0) --
(0.2cm,0) -- (0.1cm,0.2cm);}}
\newcommand{\mysquare}[1]{\tikz{\node[draw=#1,fill=#1,rectangle,minimum
width=0.2cm,minimum height=0.2cm,inner sep=0pt] at (0,0) {};}}

\colorlet{pastelgreen}{green!}
\colorlet{pastelred}{red!50!}
\colorlet{pastelyellow}{yellow!50!}
\definecolor{calpolypomonagreen}{rgb}{0.12, 0.3, 0.17}

\usepackage[strings]{underscore}
\usepackage{array}
\usepackage{soul}

\newcommand{\eg}{e.g.\@\xspace}
\newcommand{\ie}{i.e.\@\xspace}
\newcommand{\etal}{et~al.\@\xspace}

\newcommand{\platform}{\textit{CERTainty}\xspace}
\newcommand{\iris}{Iris\xspace}
\newcommand{\dnscensorship}{DNS manipulation\xspace}

\newcommand{\censor}{adversary\xspace}
\newcommand{\censors}{adversaries\xspace}
\newcommand{\insitu}{in situ\xspace}

\newcommand{\exsitu}{remote\xspace}

\newcommand{\fpr}{72.45\%\xspace}
\newcommand{\fnr}{9.70\%\xspace}
\newcommand{\bp}{blockpage\xspace}
\newcommand{\bps}{blockpages\xspace}

\newcommand{\rcode}{RCODE\xspace}
\newcommand{\metric}{heuristic\xspace}
\newcommand{\metrics}{heuristics\xspace}

\newcommand{\myparagraph}[1]{\smallskip\textbf{\textit{#1}:}\hspace{3pt}}

\newcommand{\bpcnt}{226\xspace}
\newcommand{\tlsproxycnt}{17\xspace}
\newcommand{\tlsproxycountrycnt}{52\xspace}
\newcommand{\ispcountrycnt}{26\xspace}
\newcommand{\measurementcnt}{2.93 billion\xspace}
\newcommand{\pagecnt}{31.17 million\xspace}

\newcommand{\code}[1]{\texttt{#1}}

\setcopyright{popets}
\copyrightyear{YYYY}

\acmYear{YYYY}
\acmVolume{YYYY}
\acmNumber{X}
\acmDOI{XXXXXXX.XXXXXXX}
\acmISBN{}
\acmConference{Proceedings on Privacy Enhancing Technologies}
\begin{document}

\title[CERTainty: Detecting DNS Manipulation at Scale using TLS Certificates]{CERTainty: Detecting DNS Manipulation at Scale \\using TLS Certificates}

%%%%%%%%%%%%%%%% Authors' Info %%%%%%%%%%%%%%%%%


\author{
    {\rm Elisa Tsai}$^{*}$ \quad {\rm Deepak Kumar}$^{\dagger}$ \quad {\rm Ram Sundara Raman}$^{*}$ \quad {\rm Gavin Li}$^{*}$ \quad {\rm Yael Eiger}$^{*}$ \quad {\rm Roya Ensafi}$^{*}$ 
      $^{*}${Censored Planet, University of Michigan} \hspace{2em} $^\dagger${Stanford University}\\
      $^{*}${\{eltsai, ramaks, hawklin, yaeleige, ensafi\}}@umich.edu \hspace{2em} $^\dagger${\{kumarde\}}@cs.stanford.edu
}

\renewcommand{\shortauthors}{E. Tsai, D. Kumar, and R. Sundara Raman et al.}
\begin{abstract}
  DNS manipulation is an increasingly common technique used by censors and other network adversaries to prevent users from accessing restricted Internet resources and hijack their connections. Prior work in detecting DNS manipulation relies largely on comparing DNS resolutions with trusted control results to identify inconsistencies. However, the emergence of CDNs and other cloud providers practicing content localization and load balancing leads to these heuristics being inaccurate, paving the need for more verifiable signals of DNS manipulation. In this paper, we develop a new technique, CERTainty, that utilizes the widely established TLS certificate ecosystem to accurately detect DNS manipulation, and obtain more information about the adversaries performing such manipulation.  We find that untrusted certificates, mismatching hostnames, and blockpages are powerful proxies for detecting DNS manipulation. Our results show that previous work using consistency-based heuristics is inaccurate, allowing for 72.45\% false positives in the cases detected as DNS manipulation. Further, we identify 17 commercial DNS filtering products in 52 countries, including products such as SafeDNS, SkyDNS, and Fortinet, and identify the presence of 55 ASes in 26 countries that perform ISP-level DNS manipulation. We also identify 226 new blockpage clusters that are not covered by previous research. We are integrating techniques used by CERTainty into active measurement platforms to continuously and accurately monitor DNS manipulation.

  %We work with \measurementcnt global DNS resolution data, collecting over \pagecnt HTTP(S) pages.
\end{abstract}

%\keywords{todo}

\maketitle

\section{Introduction}

%\elisa{remove the word Certainty}

Due to a lack of encryption, DNS traffic is easy to manipulate, reroute, and hijack. DNS manipulation is a common technique used by censors and other adversaries to prevent users from reaching restricted Internet resources~\cite{garcia2021large, lu2019end, doan2021measuring}. Conceptually, identifying \dnscensorship is straightforward and entails verifying the legitimacy of resolved IP addresses. However, in reality, detecting \dnscensorship on the global stage is more challenging due to website localization effects, differences in censor behaviors, and a dearth of clear signals of manipulation.

To address these challenges, prior work has proposed a myriad of detection mechanisms, most of which rely on comparing DNS resolutions and corresponding metadata collected through trusted \textit{control} DNS resolvers with those collected through \textit{test} DNS resolvers that are suspected of performing \dnscensorship~\cite{ooni, censoredplanet, iclab, gill2015characterizing, yadav2018light, trevisan2017automatic, greatgfw, chinatriplet}. Such \metrics have been deployed by longitudinal censorship measurement platforms including OONI~\cite{ooni} and Censored Planet~\cite{censoredplanet}, providing open-access data to thousands of researchers in the Internet freedom community to identify and report censorship events. Unfortunately, the rise in popularity of CDNs and cloud providers, anycast-based routing, load-balancing, DNS misconfiguration, and localization has led DNS resolutions to often be unpredictable, significantly hampering the accuracy and usefulness of the ``test metadata vs. control metadata'' strategy proposed in previous work. These issues have caused measurement platforms to, in some cases, wrongly flag instances of DNS manipulation or completely overlook instances of manipulation~\cite{yadav2018light,satellitedoc,ooni-not-censorship}. Indeed, we show in this paper that more than \fpr of the  \dnscensorship detected using state-of-the-art \metrics are false positives. Due to the far-reaching implications of censorship measurement, it is crucial that the identification of \dnscensorship is performed accurately.

In this paper, we propose a novel technique, \platform, to detect DNS manipulation by utilizing a widely adopted trust infrastructure: \textit{TLS certificates}. \platform relies on the fact that valid TLS certificates for a domain can only be issued by its owner, and DNS manipulation is performed by an in-network adversary such as an ISP that does not own the domain. \platform fetches TLS certificates from the IP addresses returned during the DNS resolution and examines the validity of these certificates for the requested domain. To do so, we equip \platform to validate certificates with a well-known root store and consider cases where TLS certificates are mismatched or untrusted. We evaluate our technique by matching responses with HTML \bps as ground truth and identify that almost all cases of certificate invalidity can be mapped back to true instances of \dnscensorship. Moreover, we use information from certificates and HTML \bps to attribute \dnscensorship and find who implements it. 

%We create and open-source \bpcnt unique DNS \bp fingerprints by clustering collected HTTP(S) pages, which can be used by the measurement community to identify and confirm the presence of DNS manipulation. We also perform TTL-limited censorship traceroute probes to identify the orgin of the TLS certificates. Using \bps and censorship traceroutes as a source of ground truth, we measure cases where HTTPS certificates are invalid or untrusted, the hostname does not match with the queried domain, or both, and identify that almost all of these cases represent true instances of \dnscensorship, showing that HTTPS certificates are an accurate proxy for \dnscensorship. Moreover, we use information from certificates and HTML \bps to confirm and attribute \dnscensorship. 

%In this paper, we argue that the research community must move away from consistency-based \metrics toward ones that provide independently verifiable signals of \dnscensorship. We build \platform, a system that uses the tandem of two such signals, the validity of HTTPS certificates and the presence of HTTP \bps, to accurately detect \dnscensorship. \platform implements robust DNS measurement strategies to extract potentially manipulated IP addresses for different domains from DNS resolvers. It then automatically performs certificate  validity checks and hostname matching on the certificate chains fetched through HTTPS, and detects \bps in HTML data. \looseness=-1

\begin{table*}[t!]
    \footnotesize
    \centering
\begin{tabular*}{2\columnwidth}{
m{3cm} 
m{2.4cm}
m{0.2cm}
m{0.4cm}
m{0.4cm}
m{0.4cm}
m{0.4cm}
m{0.4cm}
m{0.4cm}
m{0.4cm}
m{0.4cm}
m{0.4cm}
m{0.4cm}
m{0.4cm}
m{0.4cm}}
        \toprule
        \multirow{2}{2cm}{} & \multirow{2}{2.5cm}{Measurement Range} & 
        
        \multicolumn{10}{c}{\bf Consistency} & \multicolumn{3}{c}{\bf Verifiable Signals} \\
        &&&IP & HTTP & Cert & ASN & ASNa & PTR & TTL & Thres & & Cert & Page & Manual \\

        \midrule
        OONI (2012) \protect\cite{ooni} 
        & Global (>200 countries) 
        && $\bullet$ & & & $\bullet$ && $\bullet$  &&&&&$\bullet\diamond$&
        \\
        %\midrule
        %\platform $\star$
        %& Global (221 countries) 
        %&& $\bullet$ &&&&&&&&&$\bullet$& $\bullet\diamond$ &
        %\\
        
        \midrule
        Censored Planet (2020)\protect\cite{censoredplanet}
        & Global (221 countries) 
        && $\bullet$ & $\bullet$ & $\bullet$ & $\bullet$ & $\bullet$ & $\bullet$ &&&&&&
        \\
        
        \midrule
        IClab (2020)\protect\cite{iclab, gill2015characterizing} $\star$
        & Global (62 countries) 
        && $\bullet$ & &  & $\bullet$ &&&&$\bullet$&&&$\bullet\diamond$&
        \\
        
        \midrule
        Yadav \etal (2018) \cite{yadav2018light} 
        & India 
        && $\bullet$ & &  & $\bullet$ & &&&&&&&$\bullet$
        \\
        
        \midrule
        Iris (2017)\protect\cite{iris} $\star$
        & Global (151 countries) 
        && $\bullet$ & $\bullet$ & $\bullet$ & $\bullet$ & $\bullet$ & $\bullet$ & & & & $\bullet$ & &
        \\
        
        \midrule
        REMeDy (2017)\protect\cite{trevisan2017automatic} 
        & Local ISPs 
        && $\bullet$ & & & $\bullet$ & & & $\bullet$ & & & & &
        \\
        
        \midrule
        UBICA (2015)\protect\cite{aceto2015monitoring, aceto2016analyzing} 
        & Pakistan, South Korea and Italy 
        && $\bullet$ & & & & & & & & & & $\bullet$ &
        \\

        \midrule
        Verkamp \etal (2012) \protect\cite{verkamp2012inferring} 
        & Global (11 countries) 
        && $\bullet$ & & && & $\bullet$  &&&&&&
        \\
        
        \bottomrule
    \end{tabular*}
    \caption{\textbf{\dnscensorship detection \metrics}---We omit platforms that deploy country-specific measurement techniques based on the targeted DNS manipulation systems \eg China \protect\cite{greatgfw, chinatriplet, lowe2007great} and Iran \protect\cite{alexiran}.  The $\star$ symbol indicates the platforms that consider RCODE and private IPs. All the heuristics mentioned are for DNS manipulation with public IPs returned. The consistency heuristics contain control-matching IP, HTTP hash, certificate hash, AS number, AS name, PTR record, packet TTL field, and tuned threshold (for the number of domains mapping to the same IP).  The verifiable signals heuristics include the certificates (with and without SNI) fetched from the resolved IPs, web page (length and \bps $\diamond$), and manual analysis.}
    \label{tab:previous_work}
\end{table*}

We evaluate our research with previous studies that have used TLS certificates and state-of-the-art heuristics that rely on control metadata to detect \dnscensorship~\cite{iris, censoredplanet, satellite}, and find that techniques used previously are highly flawed. We discover that previous work does not properly consider the effects of hostname matching, certificate misissuance, and captive portals, leading to poor performance of TLS certificate-based detection. Moreover, we observe that the dynamic behaviors of CDN and website content localization cause \fpr of \dnscensorship cases detected using ``test vs control'' heuristics to be false positives. Moreover, these heuristics also fail to detect \fnr of the true cases of \dnscensorship identified by \platform.

We show how certificate validation and \bp fingerprint matching provide a holistic view of \dnscensorship, allowing not only detection but attribution. Globally, \platform identifies \tlsproxycnt TLS proxy vendors in \tlsproxycountrycnt countries, including products such as SafeDNS, SkyDNS, and Fortinet. We discover 7 commercial products that are deployed in more than one country and find that these products return a small pool of common tampered IP addresses across different countries, providing a good avenue for both detection as well as circumvention. \platform also detects 55 ASes in \ispcountrycnt countries with ISP-level \dnscensorship, which includes both countries with well-known DNS blocking systems (\eg Russia~\cite{decentralized} and China~\cite{chinatriplet, marczak2015analysis, greatgfw, hoang2019measuring, wang2017your}), as well as countries that have not been studied in previous research (\eg Nepal, Latvia, Poland, and Singapore). In both the commercial filtering product deployments as well as ISP-level deployments, we find a wide variety in the types of certificates and details in \bps returned. To our best knowledge, this is the first report of \dnscensorship deployment (products and ISPs) on a global scale.
%We additionally uncover instances of censorship leakage, whereby China's \dnscensorship methodology potentially impacts resolvers in 14~other countries.

Our results highlight the advantages of using established trust mechanisms such as certificate validation and \bp matching, not only in accurately detecting \dnscensorship, but also in gaining more knowledge about the \censors performing such manipulation. We have integrated \platform into Censored Planet~\cite{censoredplanet}, a remote censorship measurement platform, and data generated by \platform is actively being published and utilized by hundreds of researchers. We are also actively working on integrating our techniques into other measurement platforms such as OONI~\cite{ooni}. We hope that our techniques bring improved accuracy and rigor to the continued monitoring of \dnscensorship attempts.

\section{Background}
\label{sec:background}

%\elisa{TODO: Kill Iris in the Background. Later in Eva we should not emphasize Iris}

In this paper, we define \dnscensorship as the phenomenon where 
a network adversary --- such as a censoring authority --- manipulates DNS responses to prevent a user from accessing legitimate content for the name requested in the DNS query. \dnscensorship has been studied both in country-specific contexts~\cite{alexiran, oldgfw, chinatriplet, greatgfw, pakistan, egypt, decentralized, chaabane2014censorship, yadav2018light, bock2020detecting} and with global measurements~\cite{ooni, iclab, censoredplanet, satellite,trevisan2017automatic}, revealing its diverse and decentralized nature. Countries like Pakistan use a nonzero Response Code (\rcode), \eg NXDOMAIN, to deny access to blocked domains~\cite{pakistan}. Others, like Russia, manipulate DNS responses at the ISP level and redirect users to \bps~\cite{decentralized}. A few countries perform manipulation in a more centralized manner, deploying their national firewall at the Internet backbone, and either return private IPs~\cite{alexiran, xu2011internet} or a pool of designated IPs~\cite{greatgfw, chinatriplet}. The technological barrier to deploying \dnscensorship on a national scale has fallen, as middlebox vendors from nations with developed filtering technologies increasingly export their wares to those without them, making censorship implementation simple~\cite{deibert2012global, wagner2012push, filtermap, decentralized}.

\myparagraph{Measuring DNS Manipulation}
%\elisa{Maybe here we should explain how previous work all fall under 2 category: consistency and independent verifiability.}
%In recent years, researchers have proposed and deployed systems that launch continuous censorship measurements, and in many cases, provided such data openly for future research.
 As established in previous work \cite{ooni, iclab, censoredplanet, iris}, the \dnscensorship we aim to detect contains two scenarios: (1) the resolution is unsuccessful either because an in-path adversary drops the connection, or because poisoned or tampered DNS resolvers return nonzero \rcode for domains on the blocklist, and (2) resolved IPs do not host the requested domains \eg private IPs or IPs hosting a \bp stating that access is blocked. 

%Global measurements of \dnscensorship, and censorship in general, largely fall into two categories: \insitu and \exsitu measurements. \Insitu measurements are those conducted by a client device \textit{inside the country being studied} \eg OONI \cite{ooni} (volunteers) and IClab \cite{iclab} (VPNs). In contrast, \exsitu measurements typically use \textit{public Internet infrastructure to measure country-wide behavior}, like reflecting queries through open DNS resolvers in a given country \eg Iris \cite{iris} and Censored Planet \cite{censoredplanet}. Each measurement platform has its own merits. \Insitu measurement platforms can confidently reflect the direct experience of end-users. On the other hand, \exsitu platforms extend the coverage, continuity, and scale of measurement.

\myparagraph{DNS Manipulation Detection Heuristics}
As shown in Table~\ref{tab:previous_work}, most censorship measurement platforms such as OONI~\cite{ooni}, Censored Planet~\cite{censoredplanet}, IClab~\cite{iclab}, \iris~\cite{iris}, REMeDy~\cite{trevisan2017automatic} and UBICA~\cite{aceto2015monitoring}  incorporate a ``test vs. control`` strategy, with requests to trusted resolvers acting as control \textit{ground truth}. However, in the current Internet landscape that contains localization effects, it is challenging to ensure that such controls identify all intended resolutions. Therefore, when comparing IP addresses to control measurements is inconclusive, measurement platforms use a variety of other control-matching heuristics to determine whether DNS resolution is correct. These heuristics often fall into two categories: (1) consistency-based \metrics and (2)  verifiable signals. 

%\iris\cite{iris}, the state-of-the-art  \dnscensorship measurement system, introduced two \cats of \metrics: (1) \textit{consistency} and (2) \textit{independent verifiability}. These encapsulate previous detection \metrics, as shown in Table~\ref{tab:previous_work}. 

\myparagraph{Detecting DNS Manipulation through consistency-based \metrics}
The design philosophy of consistency-based \metrics is to confidently identify \textit{unmanipulated} DNS responses. If the IP address or \textit{any} of the other metadata matches with the corresponding metadata in the control group, the DNS response is tagged as unmanipulated. The \metrics were designed based on the insight that, behind the same domain name, there are typically shared infrastructural signals even if the exact IP address is different. For instance, Pearce \etal showed in 2017 that \metrics such as the AS number and name, HTTP content hash, HTTPS certificate hash, and PTR records serve as good consistency \metrics~\cite{iris}. Censored Planet, an active censorship measurement platform, also uses similar \metrics~\cite{censoredplanet}. 

Despite their usefulness, we show in this paper that consistency-based \metrics such as the ones introduced in Pearce \etal~\cite{iris} face a number of challenges that make them unsuitable for large-scale \dnscensorship detection (see more discussion in \S\ref{sec:eva_stat}). First, consistency-based \metrics rely on the availability of infrastructure metadata, such as AS information, which is not always accurate~\cite{poese2011ip}. Second, consistency-based \metrics result in a number of false negatives due to the fact that legitimate content and adversaries could both use the same infrastructure, such as hosting information on a CDN. Finally, consistency-based \metrics also result in a large number of false positives, since legitimate content could be hosted in different infrastructures in different regions. Because of these challenges, in this paper, we instead rely on designing verifiable signals of \dnscensorship.

%The \textit{independent verifiability \cat} uses information separate from the DNS response, including the certificates fetched from resolved IPs, contents of fetched pages, and manual verification.

\myparagraph{DNS Manipulation Detection through verifiable signals}
An alternate approach to detecting \dnscensorship is to use independent signals that can indicate whether the IP address returned during DNS resolution provides legitimate content. 
For instance, if injected or poisoned IPs redirect traffic to a \bp citing the reason for blocking, we view it as a very strong signal of \dnscensorship. Previous work has used a range of clustering techniques to identify \bps. Dalek \etal created signatures for 4 URL filter vendors in 2014~\cite{dalek2013method}. In 2014, Jones \etal utilized page length and term frequency vectors~\cite{jones2014automated} to discover \bps.  In 2020, Niaki \etal used textual similarity and HTML structure similarity to cluster potential \bps \cite{iclab}. In the same year, Sundara Raman \etal created \bp fingerprints on the application layer for more than 90 vendors and actors \cite{filtermap}. All of these previous studies manually curated the fingerprints of the \bps. Human identification remains the primary mechanism to identify the unique parts of \bps of various domains.  

Pearce \etal \cite{iris} used certificates (both with and without SNI) to identify unmanipulated DNS resolution. While they reported below-average performance in the use of certificates for detecting \dnscensorship, we find that their approach is error-prone (\S\ref{sec:eva}). We instead demonstrate in \S\ref{sec:cert} that with proper consideration of hostname-matching, certificate misissuance, and captive portals removal, the validity of certificates with domain SNI can serve as an effective detector of \dnscensorship. Using certificate validation, we not only pinpoint the signals and the corresponding actors of \dnscensorship, but also detect more covert forms of \dnscensorship where no clear signals of blocking are shown to the user, often granting the \censors plausible deniability \cite{houmansadr2011cirripede}.

\section{Data}
\label{sec:system}

We leverage open-access global DNS measurement data provided by Censored Planet~\cite{censoredplanet}. Censored Planet performs measurements to thousands of open DNS servers longitudinally, and uses consistency-based \metrics such as AS number and name, HTTP content hash, and HTTPS certificate hash to determine \dnscensorship. We propose two novel techniques for improving Censored Planet's \dnscensorship detection, both of which involve making an HTTP(S) connection to the IP addresses returned during the DNS process: 1) When a TLS Client Hello message with the appropriate Server Name is sent to resolved IP addresses, \platform checks the validity and correctness of the returned certificates, and 2) \platform clusters and identifies \bps, and determines whether the web page returned during the HTTP request matches our list of expert-curated \bp fingerprints.

\subsection{Censored Planet DNS Data}
\label{subsec:datacollection}
% Table Before Roya-styled
% \begin{table}[t]
%     %\scriptsize
%     \centering
% \begin{tabular}{ | m{10em} | m{2cm}| m{2cm} | } 
%         \hline
%         {Fetch Page Status} & {Control Pages} & {Test Pages} \\
%         \hline\hline
%         {HTTP Page Only} & 3.74\% & 7.94\%  \\
%         \hline
%         {has HTTPS Page} & 94.18\% & 83.70\%  \\
%         \hline
%         {Neither Collected} & 2.08\% & 8.35\%\\
%         \hline
%     \end{tabular}
%     \caption{\textbf{Page Fetch Result Distribution}---for control group and test group respectively}
%     \label{tab:ctrl_v_test}
% \end{table}

\begin{table}[t!]
\centering
\small
\vspace{-5 pt}
\begin{tabular}{m{12em} m{2cm} m{2cm}}
\toprule
{Fetch Page Status} & {Control Pages} & {Test Pages} \\
\toprule
{Has TLS cert and HTTP page} & 5,898 (94.18\%) & 530,170 (83.71\%)  \\
\midrule
{HTTP Page Only} & 234 (3.74\%) & 50,287 (7.94\%)  \\
\midrule
{Connection error for both HTTP and HTTPS} & 130 (2.08\%) & 52,884
(8.35\%)\\
\midrule
{Total} & 6,262 & 633,341\\
\bottomrule
\end{tabular}
\caption{\textbf{Page Fetch Result Distribution}---for control group and test group respectively.}
\label{tab:ctrl_v_test}
\end{table}

In this section, we describe the global DNS resolution data collected by Censored Planet\cite{censoredplanet}, an open-access remote website accessibility measurement platform. Note that the techniques we propose in this paper can be integrated into both \insitu (inside a country of interest, running on VPNs or volunteer machines) as well as \exsitu (utilizing open DNS resolvers in the country of interest) measurements. Since \exsitu measurements offer the advantage of increased scale and coverage, we deploy and evaluate our detection techniques on remote measurement data in this paper. 

\myparagraph{Trusted Control Resolvers}
\label{sec:sys_trusted_resolvers}
Censored Planet re-implements the consistency-based \metrics from Pearce \etal\cite{iris}, and it compares DNS resolution results collected from a set of \textit{trusted control resolvers} to the results of the test DNS resolvers. Censored Planet utilizes trusted, load-balanced, public open resolvers such as those operated by Google, Cloudflare, and UltraDNS as controls. 

\myparagraph{Test Resolvers}
Censored Planet leverages Censys~\cite{censyspaper} to locate open resolvers~\cite{durumeric2013zmap, censyspaper}, and filters these resolvers specifically to target \textit{only} ones that can be identified as infrastructure DNS resolvers by retaining resolvers with PTR records containing \code{ns[0-9]+},\newline \code{nameserver[0-9]*}, \code{.telecom} or \code{.isp.}. This criterion is important, as erroneously leveraging user-controlled resolvers in censored countries can lead to unwanted legal or government action against citizens~\cite{chinarealname1,chinarealname2,yang2017door}.
This process yields more than 25,000 open resolvers, spanning more than 200 countries.

%that we ensure do not tamper with DNS or application-level responses. We do not set up our own control resolvers due to the possibility that authoritative resolvers may block our measurement vantage point. Instead, we utilize trusted, load-balanced, public open resolvers such as those operated by Google, Cloudflare and UltraDNS as controls. 

\myparagraph{Domain Test List}
The domains tested by Censored Planet are a combination of (1) the Citizen Lab Global Test List~\cite{testlist}, which is a curated list of URLs intended to enable global censorship measurements. As of Nov 2022, the list has 1,598 domains and (2) 500 top domains from the Tranco 1M~list~\cite{tranco}, which is a list of popular domains updated daily. 

\myparagraph{Censored Planet Data Characteristics}
In this paper, we utilize Censored Planet~\cite{censoredplanet} DNS data collected twice a week from May 16, 2022, to Nov 30, 2022. Each measurement snapshot consists of sending queries to an average of 25,943 open resolvers for 2,000 domains selected as mentioned above, resulting in over 2.93 billion lines of DNS resolutions in total. Among the 2.93 billion DNS resolutions, 96.87\% succeed in getting a DNS response (i.e. do not experience a timeout), 0.006\% receive a nonzero \rcode, and 93.45\% of queries have at least one IP returned. For connection errors and nonzero \rcode, determining if the domain is accessible in a given region is a simple boolean check. However, when one or more IPs are returned, things are more complicated since many domains own myriad points of presence across the globe, hosted by various CDNs, which the control DNS resolutions fail to cover. This set of DNS resolutions is the main focus of this paper. 

\subsection{Fetching Content From Resolved IPs}

\begin{figure}[t] 
 \centering
 \includegraphics[width=0.99\columnwidth]{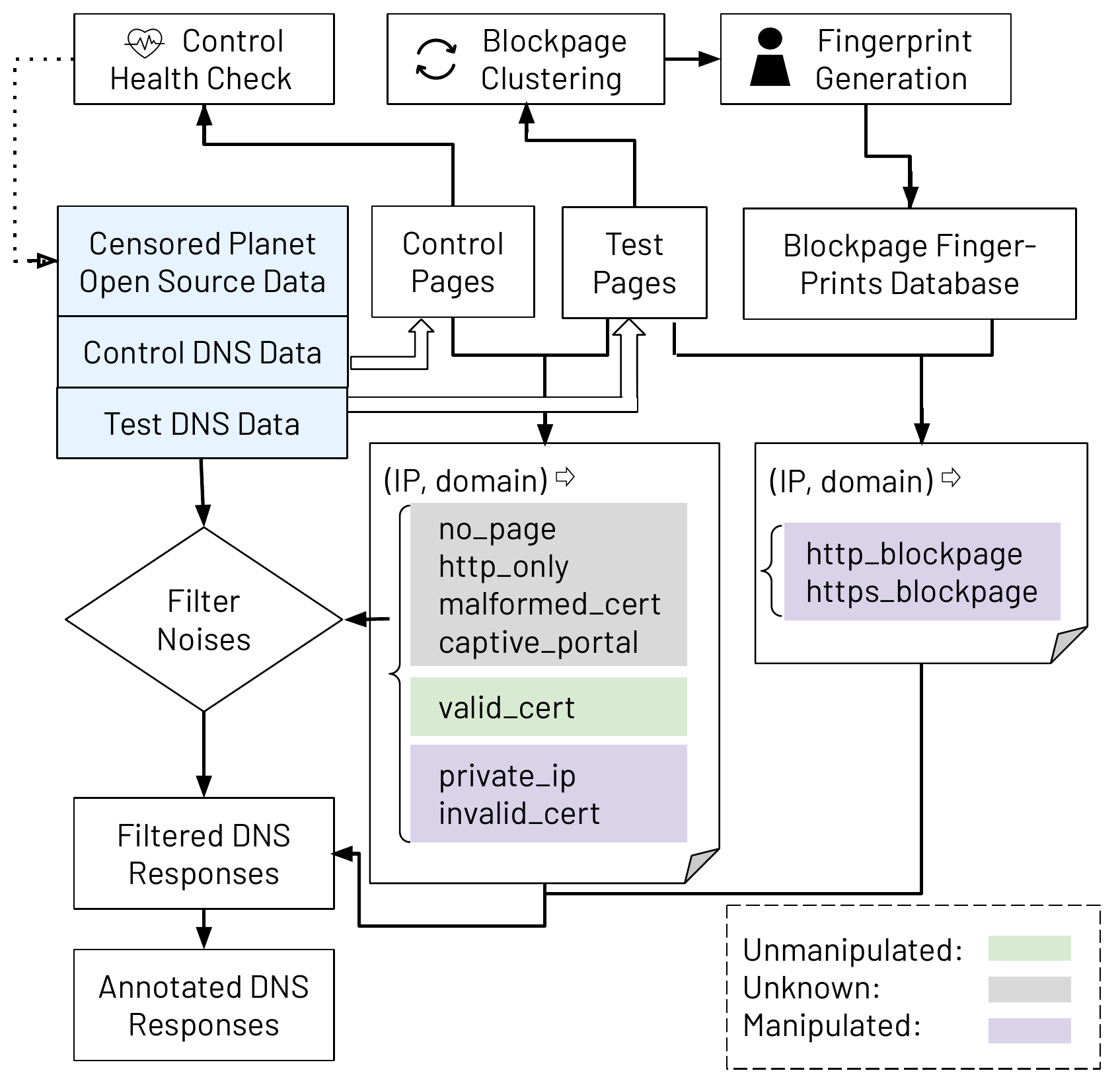}
 \caption{\textbf{The \dnscensorship detection, noise removal, and annotation pipeline of \platform}---The blue part indicates data retrieved from Censored Planet. The rest are data collected for this study.}
\label{fig:write_back}
\end{figure} 

As shown in Figure~\ref{fig:write_back}, in order to perform certificate validation and \bp matching, \platform performs HTTP(S) requests for all the DNS resolution pairs, determining if the result would appear as censorship to a user. The domain is populated into the HTTP Host Header and the SNI extension for HTTP and HTTPS requests respectively. The unencrypted HTTP header and HTML pages are used for \bp clustering to generate fingerprints. The certificate chains collected by the HTTPS requests are collected for certificate validation. We perform all page fetching measurements from a vantage point in North America. 

\myparagraph{Measurement Characteristics}
We issue \pagecnt HTTP(S) page requests for \measurementcnt DNS resolutions, over the course of 6 months. For each scan, there are on average 6,639,603 unique DNS resolution pairs. We perform HTTP (port 80) and HTTPS (port 443) page fetches to these DNS resolution pairs. As shown in Table~\ref{tab:ctrl_v_test}, we successfully connect to port 443 and obtain a certificate and an HTTP response in over 83.71\% of \code{(IP, domain)} pairs from our test cases, and obtain an HTTP page over port 80 in another 7.94\% of cases. In about 8.35\% of the cases, we see a TCP-level connection error for  both HTTP (port 80) and HTTPS (port 443) requests. 

The HTTP(S) connection errors \textit{could} be a signal of \dnscensorship that requires further investigation. For example, if TCP resets are observed, it could be the \censors resetting TCP connection for blocked domains (although this is unlikely since our measurement infrastructure is within a University network). More likely, these are cases where domains are not active on port 80 or port 443, or these domains geoblock requests from our measurement infrastructure. We discuss the connection errors in \S\ref{sec:discuss}, and provide a case study in \S\ref{sec:conerr}. %In Appendix~\ref{sec:rcode_case_study} we demonstrate a case study of Russia AS20485 where both requests failed but a \bp is served on port 444.

\subsection{Blockpage Fingerprint Generation}
\begin{figure}[t] 
 \centering
 \includegraphics[width=0.99\columnwidth]{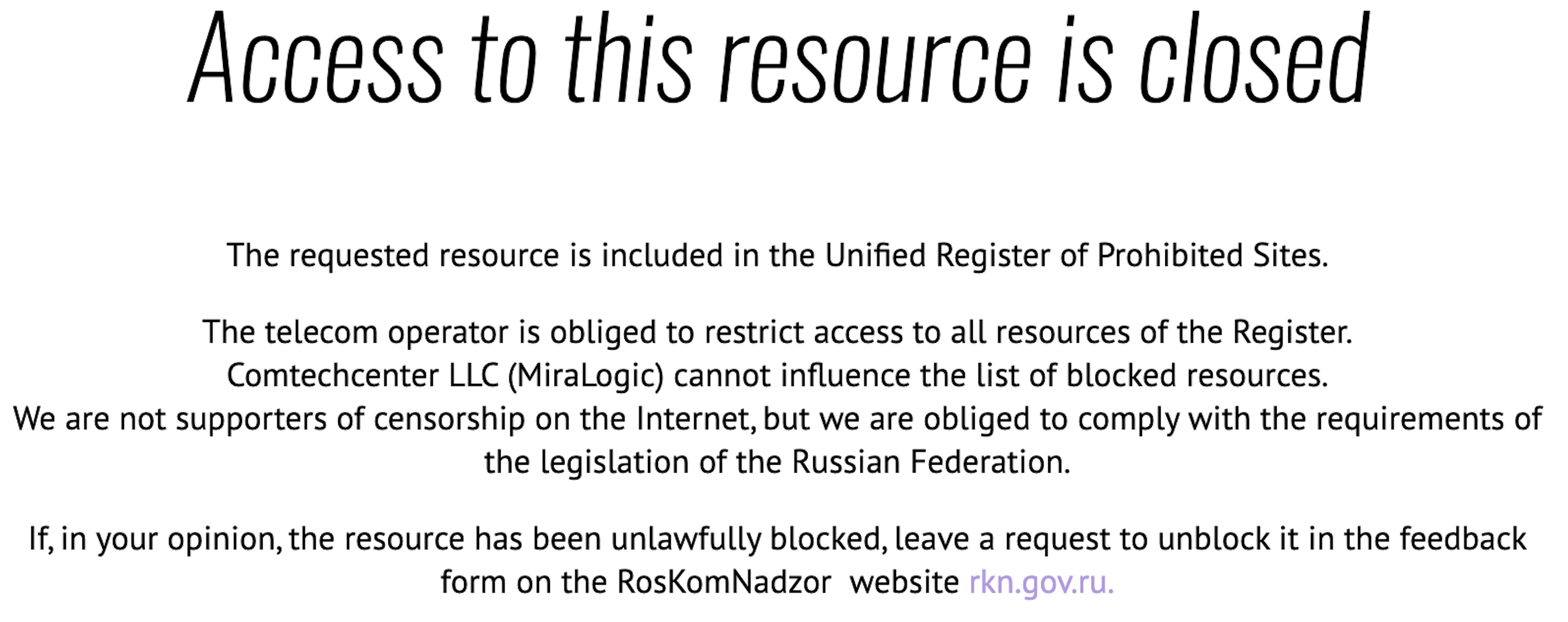}
 \caption{\textbf{Example of \bps \platform detects} -- A translated \bp from MiraLogic, a Russian telecom company}
 %stating that "We are not supporters of censorship, but we are obliged to comply with the legislation of the Russian Federation."}
\label{fig:rubp}
\end{figure}

In order to capture signals of overt censorship where a \bp is served, \platform fetches HTTP response headers and HTML pages from the IPs returned by the control resolvers and test resolver. Previous work has investigated different methods to identify \bps, utilizing clustering techniques based on the similarity of page length~\cite{iclab, jones2014automated}, term frequency vectors~\cite{jones2014automated}, HTML structure~\cite{iclab} and the screenshot of the returned pages~\cite{filtermap}. 

All of these \bp clustering techniques are followed by a step to manually create \bp fingerprints and appropriate labels for them. Constructing meaningful fingerprints manually is a widely accepted practice of \bp detection. Tedious as it seems, these \bp fingerprints provide valuable insights for the research community to track the scope, scale, and evolution of content-based censorship. 

In this paper, we integrate publicly available \bp fingerprints~\cite{cpbp} generated previously from HTTP and HTTPS connection interference data~\cite{filtermap}, and complement it with \bpcnt new \bp fingerprints generated from Censored Planet DNS data. We observe that clustering the pages in the HTTP response based on page length and HTML structure works the best. Figure~\ref{fig:rubp} is an example of ISP-level \bp discovered by our semi-automatic \bp detection. In total, \ispcountrycnt countries' ISP-level \bp are discovered, including countries that are not covered by previous research \eg Nepal, Latvia, Poland, and Singapore. 

% Table Before Roya-styled
% \begin{table}[ht!]
% \centering
% \begin{tabular}{|p{9mm} p{8mm} p{8mm} p{3mm} p{12mm} p{10mm} p{8mm}|} 
%  \hline
%  {Category} & {Product} & {National} & {ISP} & {Corporation} & {Unknown} & {General} \\ 
%  \hline
%  Count & 26 & 92 & 38 & 14 & 15 & 30 \\ 
%  \hline
% \end{tabular}
% \caption{\textbf{Blockpage fingerprint distribution}: The number of unique \bp fingerprints under different categories.}
% \label{table:fp_dist}
% \end{table}

\begin{table}[t!]
\centering
\small
\vspace{-5 pt}
\begin{tabular}{p{9mm} p{8mm} p{8mm} p{3mm} p{12mm} p{10mm} p{8mm}}
\toprule
{Category} & {Product} & {National} & {ISP} & {Corporation} & {Unknown} & {General} \\ 
\midrule
Count & 29 & 92 & 46 & 14 & 15 & 30 \\ 
\bottomrule
\end{tabular}
\caption{\textbf{Blockpage fingerprint distribution}: The number of unique \bp fingerprints under different categories.}
\label{table:fp_dist}
\end{table}

We manually verify each blockpage cluster in order to remove false positives. The presence of \bps is relatively stable in our dataset. For over 8 months' time (from March 2022 to Nov 2022), only 21 new potential clusters are observed. Therefore future manual efforts for generating \bp fingerprints are manageable. We craft the \bp fingerprints into 6 categories as shown in Table \ref{table:fp_dist}, following the convention set by previous work~\cite{filtermap}:
\begin{enumerate}
    \item \textit{Commercial product}: commercial middleboxes \eg\newline OpenDNS~\cite{contreras2022}, NextDNS~\cite{nextdns}, and OneDNS~\cite{onedns}. 
    \item  \textit{National-level \dnscensorship}: \eg Indonesia \bps that contains ``Internet Positif (Positive Internet)``.
    \item \textit{ISP-specific \bps}: \bps which specify the ISPs who configure the \bps \eg Fig~\ref{fig:rubp}.
    \item \textit{Corporational or institutional \dnscensorship}: \eg blocking implemented by companies and universities. 
    \item \textit{Unknown}: \bps that we do not have enough information to attribute the deployer. Each such fingerprint is annotated with the country of origin.
    \item \textit{General}:  \eg ``This Site Has Been Blocked`` in the title.
    
\end{enumerate}

Censored Planet has already incorporated HTTP(S) page fetch, certificate validation, and \bp fingerprint matching into their weekly measurement process. Therefore, the \bp fingerprints are also used to conduct health checks for control resolvers, allowing future control resolver list expansion. The annotated fingerprints are open-sourced for the community. We utilize \bp fingerprints to serve as ground truth in evaluating certificate validation in \S\ref{sec:cert}.

\subsection{Noise Removal}
\label{sec:remove_noise}
Censored Planet issues about 50 million DNS queries to more than 25,000 open resolvers across the globe. We add an extra filtering stage to exclude resolvers and IPs with erroneous behaviors. For example, if a resolver is returning erroneous responses for all of the queried domains, it is highly unlikely because of \dnscensorship. Instead, the cause is possibly misconfiguration or misguided NATs and firewalls~\cite{kreibich2010netalyzr}. Therefore, we exclude DNS responses from resolvers whose DNS responses either only contain timeouts, or have a nonzero \rcode, an empty list of IP addresses, private IP addresses, or the same set of IP addresses (possible captive portals) for \textit{all} the queried domains.

Previous platforms proposed by Sundara Raman \etal~\cite{censoredplanet} and Pearce \etal~\cite{iris} ignore connection errors and nonzero \rcode responses entirely since many are not due to \dnscensorship. We deploy a more conservative filtering strategy to find signals of \dnscensorship in these responses. In \S\ref{sec:rcode_case_study}, we show how proper filtering can help us capture signals of \dnscensorship within the NXDOMAIN \rcode.

\subsection{Ethics}
\label{sec:ethics}
In this paper, we use DNS measurement data collected by Censored Planet, which follows best practices in selecting measurement vantage points and conducting measurements~\cite{censoredplanet}. Censored Planet only selects DNS resolvers belonging to the Internet infrastructure such as nameservers for performing measurements. As highlighted by previous work, this is an attempt to eliminate any use of resolvers or forwarders owned by individuals~\cite{iris, censoredplanet, pearce2017augur, vandersloot2018quack, durumeric2013zmap}. This step significantly minimizes the risk, because the risk posed to administrators with more skills and resources to understand the traffic is lower than the risk posed to end users. We also set up WHOIS records and a web page served from port 80 of our measurement machine that indicates that our HTTP and HTTPS measurements are part of a research project and offer administrators the option to opt out of our scanning. We did not receive any inquiries or complaints over the period of 6 months.

\section{Using Certificate Validity to Measure \dnscensorship}
\label{sec:cert}

Prior work in DNS manipulation detection has not incorporated TLS certificate validity into \dnscensorship detection properly~\cite{iris}. However, the presence of a valid certificate (i.e., one trusted by a known root store and containing the correct hostname) is a strong signal that the application-layer connection to a server (HTTPS) is legitimate. In this section, we examine how we can use certificate validation as a proxy to evaluate the presence of \dnscensorship.

For the scope of this study, we consider a certificate to be valid if two criteria are met: (1) the certificate chains to a trusted root in the Mozilla NSS Root Store (used by Mozilla Firefox) by OpenSSL~\cite{openssl}, and (2) the hostname in the certificate (either in the common name or the subject alternative name) matches the domain we are attempting to reach, following the rules as specified in RFC 6125 \cite{certhostname}.  We note that \platform regards expired certificates as invalid, as we are using OpenSSL to verify the chain. Our approach is similar to the one followed by a browser attempting to validate the authenticity of a domain. At a high level, we consider any connection that returns a valid certificate to be unmanipulated and use other signals (e.g., \bp fingerprints) to link certificate invalidity to \dnscensorship. We utilize 662 \bp fingerprints, from both publicly available \bps~\cite{filtermap} and \bps \platform detected.

We consider four distinct cases in certificate validation when identifying \dnscensorship where the certificate we obtain from control resolutions are valid, and one case where the certificates we obtain from control resolutions are invalid. %, and use the HTTP response status code to classify and explore them. 

Our analysis is based on 12 snapshots over the course of 6 months. We do not identify significant differences among the snapshots----we identify 8 new invalid certificate issuers and 13 new \bp fingerprints in 6 months. Therefore, the analysis in this paper is based on one snapshot in Nov 2022. We discuss the longitudinal aspect of \dnscensorship in \S\ref{sec:discuss}.

\subsection{Case 1: Valid Certificates}
We view the presence of a valid certificate for the requested domain as a strong signal that the IP address is \textit{not} manipulated. Indeed, we note that none of the HTML pages returned with a valid certificate match a known \bp fingerprint. We receive a handful (5.34\%, 315 out of 5,898 total requests) of 403 error pages from CDNs due to cases such as geoblocking, where HTTP requests from our measurement infrastructure are blocked due to their location. However, the certificate for the requested domain is still valid in these cases, suggesting that certificate validation can help to eliminate the effect of geoblocking based on the vantage point chosen for measurements.

\subsection{Case 2: Untrusted Certificate With Matched Hostname}
\label{sec:untrusted_cert_matching}
\begin{figure}[!t]
 \centering
    \begin{tabular}{ p{8cm} }
    \includegraphics[scale=0.5]{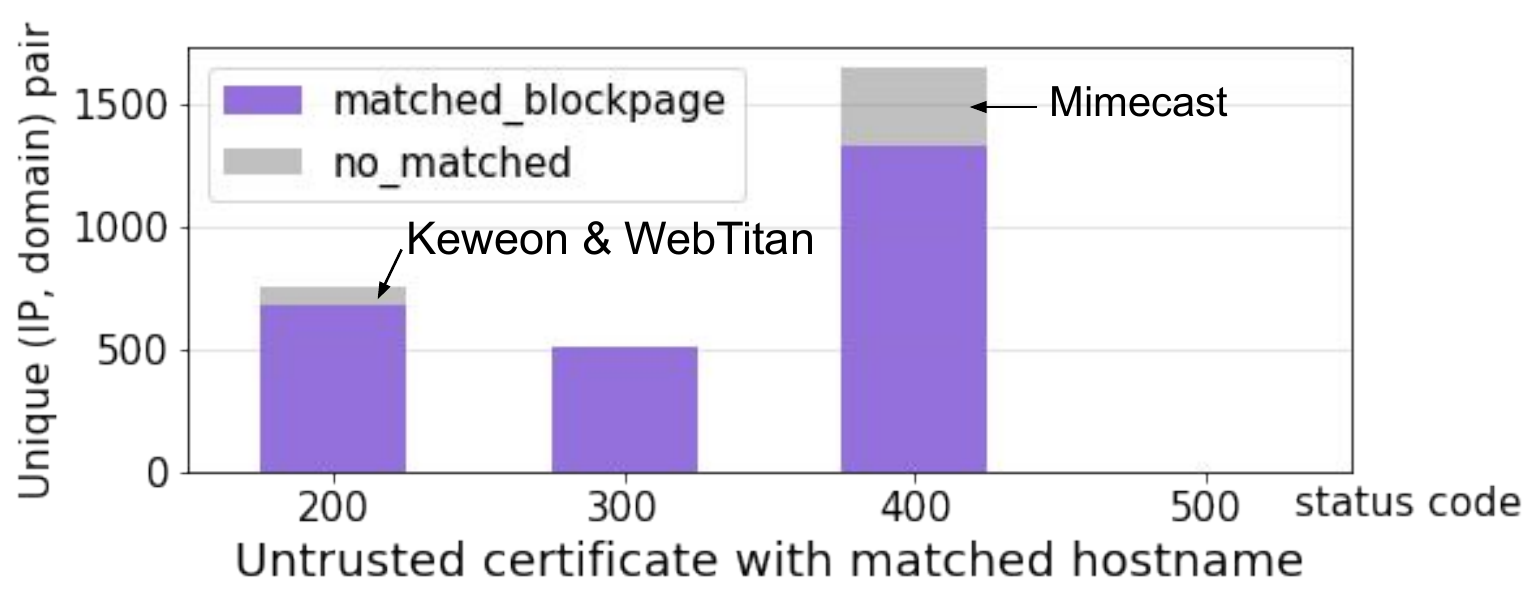}\\
    \includegraphics[scale=0.5]{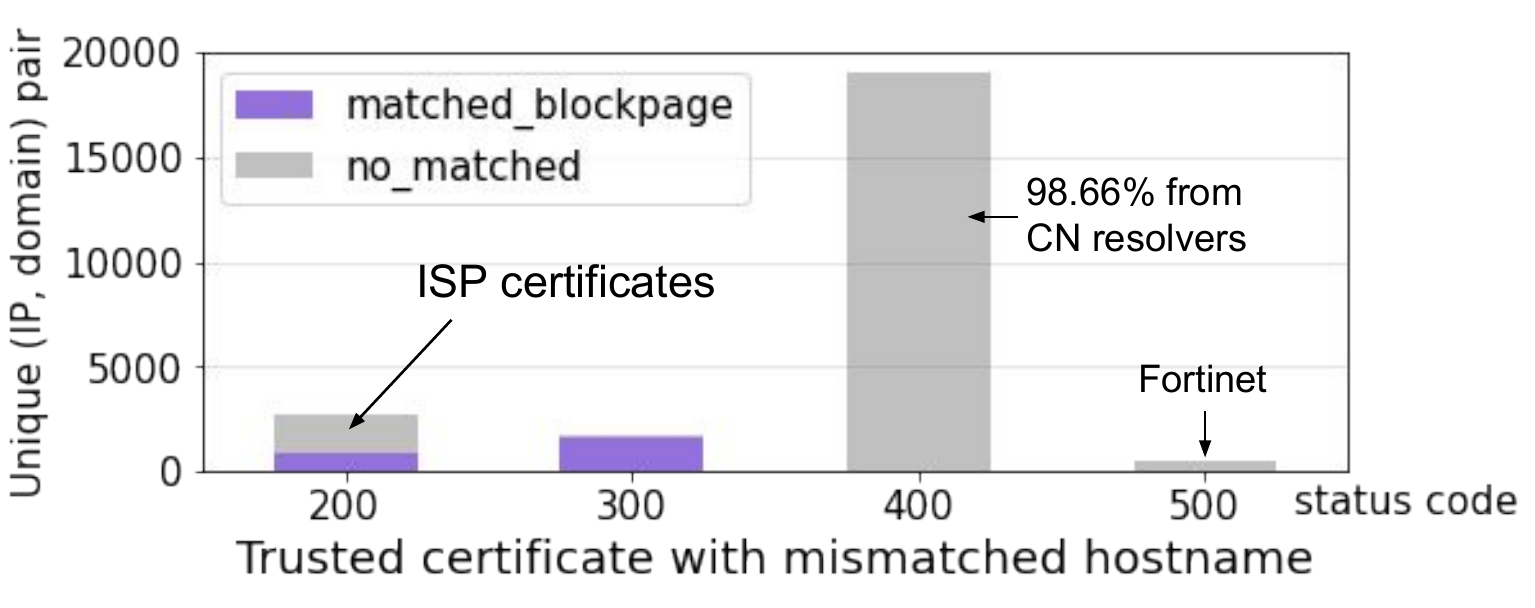} \\
    \includegraphics[scale=0.5]{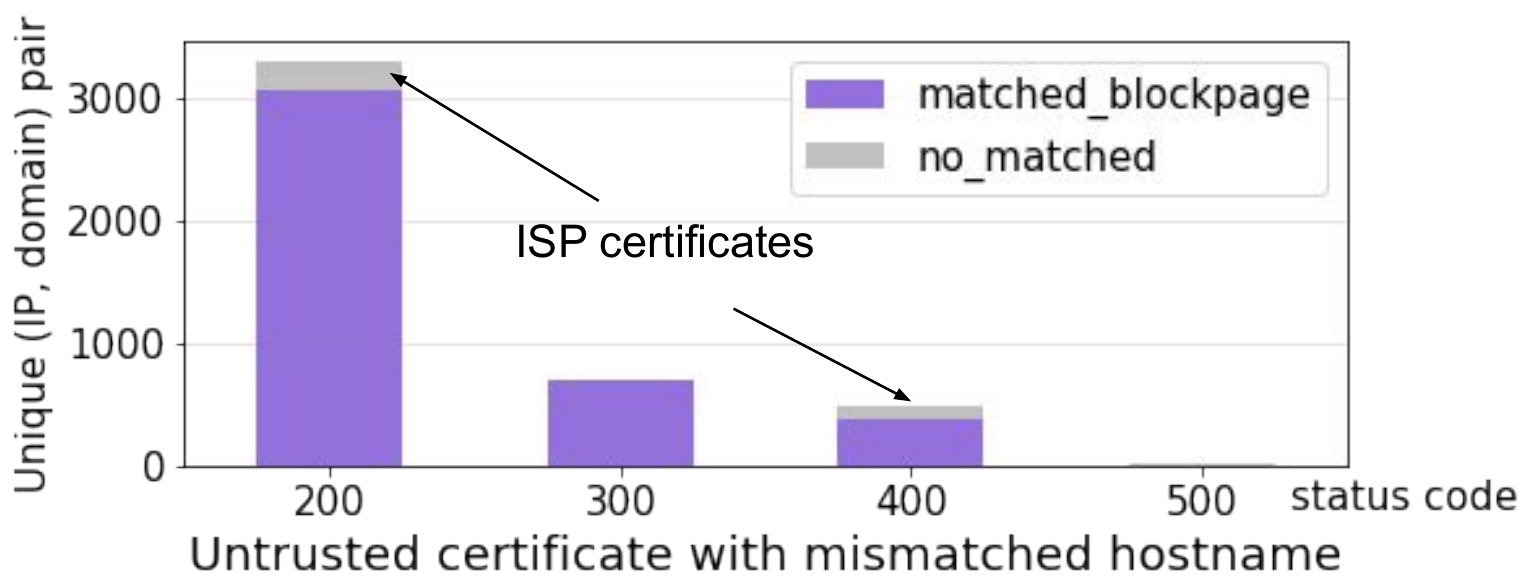}\\
    \end{tabular}
    \caption{\textbf{Blockpage fingerprint matching and control certificate matching for HTTPS responses with invalid certificate}---The purple part denotes the certificates with an identified \bp. The grey part denotes the proportion of corresponding HTML pages that do not match any known \bp fingerprint.
    IPs of captive portals and invalid certificates that match with misissued control are removed as described in Sec. \protect\ref{sec:remove_noise} and Sec. \protect\ref{sec:cert_misissuance}.}
    \label{fig:match_untrusted_cert}
\end{figure}

% \elisa{do we want to change it to more color blind version? https://www.color-blindness.com/coblis-color-blindness-simulator/ TODO change to bar chart}
If an untrusted certificate is returned with a matching hostname for a request, we mark the request as potential \dnscensorship. To confirm this categorization, we check our \bp fingerprints against the pages returned during the HTTP request. As shown in Figure~\ref{fig:match_untrusted_cert}-(1), we observe that 86.25\% (2,521 out of 2,923) of the untrusted certificates with a matching hostname come with an identified \bp. For the other 13.75\%, only 3 TLS product vendors are identified using the certificate issuer field. Keweon~\cite{keweon} and WebTitan~\cite{webtitan} return an empty 200 OK HTTP page along with the certificates, making it impossible to craft a \bp fingerprint for them. Mimecast~\cite{mimecast} returns a general 404 error page. This shows that information extracted from certificates can be critically informative when no \bps are presented. Certificates with untrusted root and matching hostname are strong signals of TLS proxies. Notably, two vendors---SkyDNS and SafeDNS---return pages with 451 status code (``Unavailable For Legal Reasons``), indicating that they are used by ISPs or governments. In total, we discover 12 such TLS proxy vendors and report our results in \S\ref{sec:res_prod}.

\subsection{Case 3: Trusted Certificate With Mismatched Hostname}
\label{sec:trusted_cert_wrong_hostname}
When a trusted certificate is returned with a mismatched hostname, we consider this to be a potential sign of \dnscensorship. Exploring these cases, we observe this behavior to be largely driven by ISP-level blocking. Of the requests made in this category, 10.48\% (2,518 out of 24,029) of them match a \bp fingerprint, as shown in Figure~\ref{fig:match_untrusted_cert}-(2). For requests that return 400+ status codes, 98.66\% (18,825 out of 19,079) of these requests are returned by Chinese open resolvers, and those IPs typically belong to large entities like Facebook (66.30\%, 12,481 out of 18,825), Twitter (29.10\% 5,478 out of 18,825), Cloudflare (3.36\% 632 out of 18,476), and other blocked CDN services \eg Fastly and Akamai (less than 0.08\%). Our observations align with prior China-focused studies~\cite{chinatriplet, marczak2015analysis, greatgfw, hoang2019measuring, wang2017your} that suggest China's national Firewall (the GFW) returns IP addresses of large US-based companies to DNS queries of blocked content. The rest are mostly from Canadian Shield~\cite{cira}, a Canadian TLS middlebox vendor. For the 1.18\% (543) requests that return a 500+ status code, we observe that 34.62\% are returned by Chinese resolvers and point to the IP address of a large entity mentioned above. The remaining 62.98\% certificates are issued by Fortinet, a well-known middlebox product vendor.  

For requests that return a 200 status code, 34.28\% (917 out of 2,675) of the returned webpages match a known \bp fingerprint. We manually investigate the 65.72\% of 200 status code webpages that we did not identify as \bps, as well as the certificates that we collected for these requests. We identify 88.22\%  are ISP-issued certificates coming with ``\code{blockpage}``, ``\code{allownet}`` or ``\code{illegal}`` in the certificate common name. For example, we see a certificate signed by \code{``illegal.mdes.go.th``} without meaningful page content, which is the Ministry of Digital Economy and Society of Thailand. This again proves that \bp information alone is not enough for \dnscensorship detection. More ISPs that return informative certificates without \bps can be found later in Table~\ref{tab:isp_res}. The rest are instances of phishing, where we see that traffic is diverted to advertisement websites. Finally, for cases where a 300 status code is returned, we observe a large number of cases where domains have misconfigured TLS certificates and discuss these further in \S\ref{sec:cert_misissuance}.

\subsection{Case 4: Untrusted Certificate With Mismatched Hostname}
\label{sec:untrusted_cert_wrong_hostname}
An untrusted certificate with a wrong hostname is a very strong signal that the returned IPs do not host the requested domain, and is therefore a potential signal for \dnscensorship. We observe 92.31\% (4,167 out of 4,514) of the pages match with a \bp fingerprint. Upon further manual investigation for the 2.57\% of unidentified pages, we find that this category of certificate likely originates from a misconfiguration. Certificates with the common name \eg ``\code{testexp}``, ``\code{test}`` and ``\code{Plesk}`` are returned with blank pages. Only 9 IPs in our whole dataset host such certificates. For the general 400+ error pages, we see certificates issued by ISPs in a few countries \eg Singapore, Columbia, and Russia. The information in the certificate (\eg common name \code{``*.block.msm.ru``}) highlights these are cases of \dnscensorship even when there is no explicit blockpage. 

\noindent \textbf{Summary:} Our results give us strong confidence that certificate validation is an effective proxy to detect \dnscensorship. It provides a venue to perform quick automated detection of DNS manipulation, reveals critical information when the middleboxes choose not to return \bps, and can even help us discover covert \dnscensorship (more in \S\ref{sec:res_covert}).

\subsection{Case 5: Invalid Control Certificate (Misissuance)}
\label{sec:cert_misissuance}
In order to use certificate validity as a proxy for detecting \dnscensorship, we need to account for certificates that would be invalid even in a control setting, as invalid certificates are common on the Internet~\cite{ukrop2019will, kumar2018tracking, akhawe2013here, acer2017wild}. These ``control certificates'' serve as ground truth for the case where the certificates are invalid because of deployment errors from domain administrators.  

We see 1.2\% (72 out of 5,898) invalid certificates among all unique control certificates collected. This means the certificates for those domains are not issued correctly by default and the presence of invalid certificates for those domains is not necessarily a signal of \dnscensorship. We fetch TLS certificates regardless of the website's HTTP-based by default. We note that the number of domains with misconfigured TLS is much lower than what previous studies have found~\cite{iris}, possibly due to increased HTTPS adoption. %Fig.~\ref{fig:invalid_ctrl_cert} contains 3 examples of the invalid control certificates.  
We see a good proportion of .mil domains failing certificate validation such as \code{www.dtic.mil}. The US military (DoD) websites utilize Federal PKI, which is not trusted by most root stores~\cite{govcert}. Other misconfigured domains (\eg \code{www.freeexpression.org} and \code{www.kcna.kp}) either have invalid certificates because of mismatching hostname, or untrusted root CA. We do not consider cases where domains have invalid control certificates as a signal of \dnscensorship.

%\begin{figure}[ht]
%\centering
    %\includegraphics[scale=0.49]{figure/iris_fn_bar.pdf}
    %\caption{\textbf{Distribution of matching IP metadata for manipulated DNS response omitted by the consistency \metrics}---The resolved IPs and/or their metadata match with the control pool, but host either an invalid certificate for the queried domain, or a \bp.}
%    \label{fig:iris_fn}
%\end{figure}

\begin{table}[t]
    %\scriptsize
    \small
    \centering
    \begin{tabular}{m{8em}  m{2em} m{2em} m{3em} m{3.8em} m{3em}}
    \toprule
    {Matched Heuristics} & {HTTP hash} & {Cert hash} & {ASN} & {AS name} & {CDN} \\ 
     \midrule
    {Count} & {372} & {460} & {10,388} & {10,384} & {11,937} \\ 
    \midrule
    {Percentage} & {3.12\%} & {$3.85\%$} & {$ 87.02\%$} & {$ 86.99\%$} & {$ 100.00\%$} \\     
    \bottomrule
    \end{tabular}
    \caption{\textbf{False negatives introduced by consistency \metrics}---The table shows the number and percentage of total cases where each consistency \metric shows a match between test and control experiments, but our technique indicates \dnscensorship due to an invalid certificate or presence of \bp.}
    \label{tab:iris_fn}
\end{table}

\subsection{Origin of Invalid Certificates}
% However, since we are issuing HTTPS requests from a university network, where no censorship is implemented. 
In this paper, we use the validity of certificates as a proxy to detect DNS manipulation. Theoretically, an adversary such as a censor that is in-path can inject invalid certificates by inspecting the HTTPS requests \eg the SNI field in Client Hello. Thus, there is the possibility that we are misclassifying network manipulation at the TLS layer as \dnscensorship. However, note that this is unlikely since we are issuing HTTPS requests from a university network where no censorship is implemented instead of sending requests from inside a censored country where censorship can happen on different network stacks. 

Nevertheless, to confirm that the certificates we receive are indeed originating from the IP addresses received during DNS resolution, we perform TTL-limited traceroute tests for \code{(IP, domain)} pairs with invalid certificates, using methods developed in previous work~\cite{raman2022network}. We perform two TLS Hello requests for the control domain (\code{exmaple.com}) and the target domain, sending the requests with incrementing TTL values. Then we compare the control traceroute and test traceroute to determine where in the network the TLS response is originating from. 

The results confirm our hypothesis. In all cases, we observe that the traceroute terminates in the same network (/24 subnet) as the endpoint IP address. Indeed,  93.24\% (5,729/6,144) of traceroutes end within \code{+-1} of the hop where the control traceroutes end. Therefore, we are confident that the certificates are returned by the IPs obtained during DNS resolution.

%\subsection{Annotating the DNS Resolution Data}

%\platform uses the validity of the certificates and \bp matching information to annotate the DNS responses collected. It tags every unique \code{(IP, domain)} pair it obtains as responses to DNS measurements as \code{\textbf{unmanipulated}}, \code{\textbf{manipulated}}, or \code{unknown}, as shown in Figure~\ref{fig:write_back}. The \code{\textbf{unmanipulated}} cases include obtaining a valid cert or the case where the domain does not implement TLS correctly (based on our control certificate). The \code{\textbf{manipulated}} cases include observing private IPs, a HTTP blockpage, or an invalid cert that does not match any of the control certificates. We consider any case where we obtain only an HTTP response that does not show a blockpage, or ones with failed TCP connections to be \code{unknown}. 
\section{Evaluation}
\label{sec:eva}

In this section, we assess the effectiveness of our method for identifying \dnscensorship using certificate validation and blockpage matching. As a baseline, we compare our technique directly against several \metrics proposed by the current state-of-the-art in two categories: (1) \textit{verifiable signals} and (2) \textit{consistency \metrics}, which compare potentially censored response data (e.g., IP, ASN, HTTP page hash) to trusted control responses.

\begin{figure}[t] 
 \centering
 \includegraphics[width=0.99\columnwidth]{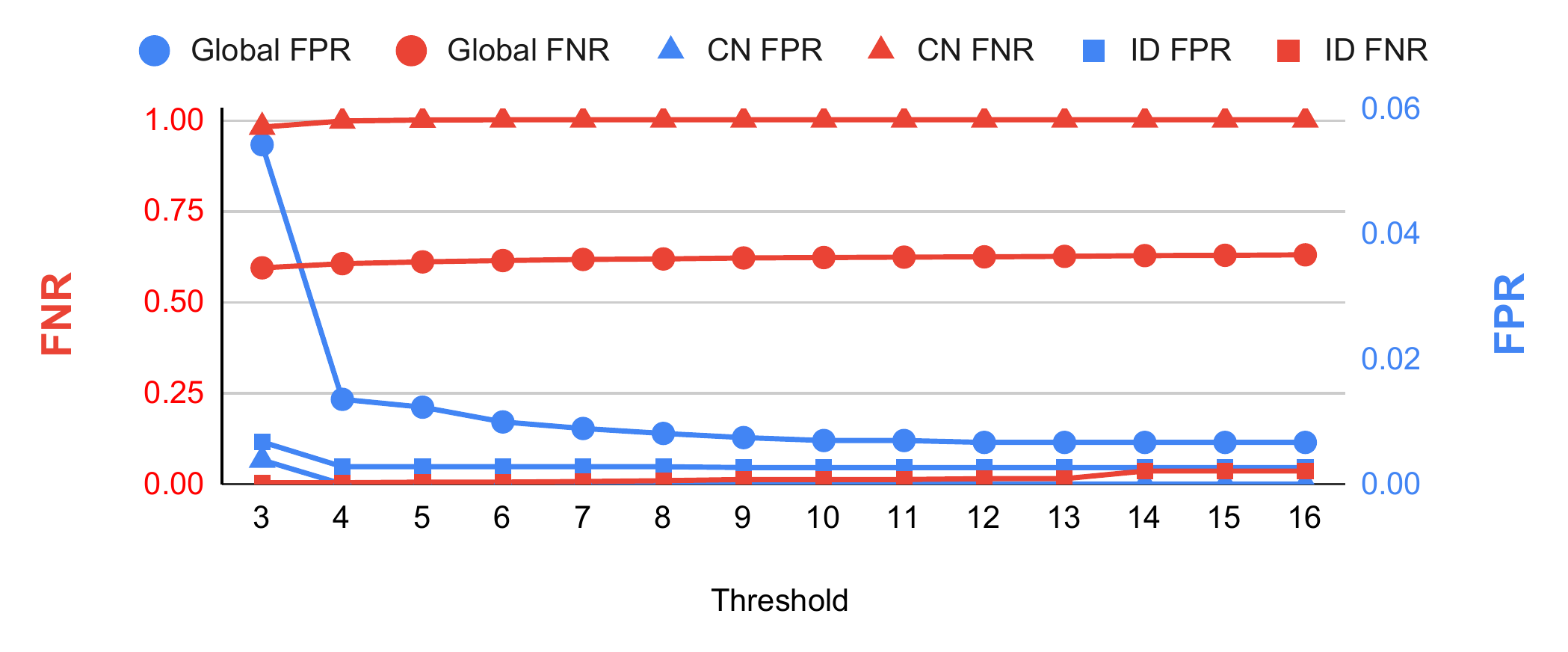}
 \caption{\textbf{The Efficacy of the AS threshold \metric}---Generated by comparing statistical AS thresholding to the results of certificate validation and \bp matching on the global scale, China and Indonesia, respectively.}
\label{fig:thres_res}
\end{figure}

\subsection{Previous Verifiable Signal Heuristics}
Some prior work leverage \textit{verifiable signals} such as certificate validation~\cite{iris}, returned pages~\cite{yadav2018light, iclab, ooni}, and manual analysis~\cite{yadav2018light}. We demonstrate how previous certificate validation is error-prone, and how our \bp fingerprints enrich the body of knowledge for censorship detection.
 
\myparagraph{Certificate Validation}
Pearce \etal's~\cite{iris} \iris technique established in 2017 checks whether the returned certificates for servers that support HTTPS are browser-trusted. For requests with SNI, the technique checks if they are for the correct IP addresses. However, in retrospect, this is no different from control IP matching. Moreover, the HTTPS ecosystem has changed a lot since 2017. In our data, only 0.12\% of certificates have IP addresses in their common name or SAN, therefore this method is not applicable anymore.

\begin{table*}
\centering
\small

\begin{tabular*}{1.99\columnwidth}{
 m{2cm}
 m{2cm} 
 m{1cm}
 m{1cm}
 m{1cm}
 m{2cm}
 m{2cm} 
 m{1cm}
 m{1cm} 
} 
\toprule
\multicolumn{5}{c}{No control hueristic matching}                                                                                                                                                                                                                                                                             & \multicolumn{4}{c}{At least one control hueristic matching}                                                                                                                                                                                                                                                                                \\ 
\midrule
\multicolumn{1}{c}{Comparison}                                    & \multicolumn{1}{c}{\platform~Result}                  & \multicolumn{1}{c}{Count}                                      & \multicolumn{1}{c}{Percentage}                                                                 && \multicolumn{1}{c}{Comparison}                                    & \multicolumn{1}{c}{\platform~Result}                  & \multicolumn{1}{c}{Count}                                          & \multicolumn{1}{c}{Percentage}                                                                   \\ 
\midrule
\begin{tabular}[c]{@{}l@{}}Same with\\\platform\end{tabular}       & \begin{tabular}[c]{@{}l@{}}Invalid Cert\\HTTP Blockpage\end{tabular} & \begin{tabular}[c]{@{}l@{}}$95,624$\\$15,492$\end{tabular} & \begin{tabular}[c]{@{}l@{}}$13.98\%$\\$2.27\%$\end{tabular} && 
\begin{tabular}[c]{@{}l@{}}Contradict with\\\platform\end{tabular} & \begin{tabular}[c]{@{}l@{}}Invalid Cert\\HTTP Blockpage\end{tabular}   & \begin{tabular}[c]{@{}l@{}}$11,097$\\$840$\end{tabular}        & \begin{tabular}[c]{@{}l@{}}$0.13\%$\\$0.01\%$\end{tabular}   \\
\midrule
\begin{tabular}[c]{@{}l@{}}Contradict with\\\platform\end{tabular} & \begin{tabular}[c]{@{}l@{}}Valid Cert\end{tabular}   & \begin{tabular}[c]{@{}l@{}}$495,532$\end{tabular} & \begin{tabular}[c]{@{}l@{}}$72.45\%$\end{tabular} &&
\begin{tabular}[c]{@{}l@{}}Same with\\\platform\end{tabular}       & \begin{tabular}[c]{@{}l@{}}Valid Cert\end{tabular} & \begin{tabular}[c]{@{}l@{}}$7,529,487$\end{tabular} & \begin{tabular}[c]{@{}l@{}}$88.85\%$\end{tabular}  \\
\midrule
\begin{tabular}[c]{@{}l@{}}Unconfirmed by\\\platform\end{tabular}  & \begin{tabular}[c]{@{}l@{}}HTTP Only\\Connection Error\\Malformed Cert\end{tabular}  & \begin{tabular}[c]{@{}l@{}}$33,592$\\$38,407$\\$5,275$\end{tabular} & \begin{tabular}[c]{@{}l@{}}$4.91\%$\\$5.61\%$\\$0.77\%$\end{tabular}  && \begin{tabular}[c]{@{}l@{}}Unconfirmed by\\\platform\end{tabular}  & \begin{tabular}[c]{@{}l@{}}HTTP Only\\Connection Error\\Malformed Cert\end{tabular}  & \begin{tabular}[c]{@{}l@{}}$186,627$\\$551,179$\\$194,390$\end{tabular}   & \begin{tabular}[c]{@{}l@{}}$2.20\%$\\$6.50\%$\\$2.29\%$\end{tabular}   \\
\bottomrule
\end{tabular*}
\caption{\textbf{Detection result comparison between consistency-based \metrics and \platform}---We view the presence of a valid certificate as a strong signal of correct DNS resolution. Invalid certificates that do not match with control, as well as the presence of \bps are strong signals of \dnscensorship. We use "malformed cert" to denote the invalid certificates that match with invalid control certificate. The consistency-based \metrics include AS number and name, HTTP hash, certificate hash, and PTR lookups.}
    \label{tab:anomaly_comp}
\end{table*}

For requests without SNI, \iris checks if the returned certificate contains the right domain name. However, many CDNs would return a general CDN certificate when no domain is specified. Our primary analysis shows that among all the DNS resolution pairs that return a valid certificate when queried with the domain as SNI, 63.89\% return general CDN certificates with mismatched hostnames, or no certificates at all when requested without SNI. This indicates that the certificate metrics proposed in previous work introduce FPs into the system. Moreover, \iris reported poor performance (40\% to 70\% accuracy) in using certificates to detect correct DNS resolution, attributing the performance issue to widespread misconfiguration of TLS servers. Since 2018, more servers have adopted TLS, and hence we find certificate validation to be more useful~\cite{httpsadoption}. 

%However, \iris reported poor performance (40\% to 70\% accuracy) in using certificates to detect unmanipulated DNS resolution, attributing the performance issue to widespread misconfiguration of TLS servers. Since 2018, more servers have adopted TLS, and hence we find certificate validation to be more useful. 

%Moreover, no prior work has taken into account \textit{hostname validation}, which is an important aspect of ensuring certificate validity. 

\myparagraph{Page information}
Previous work has incorporated information extracted from the page fetched from resolved IPs, either using the page length \cite{aceto2015monitoring} or identifying \bps \cite{iclab, ooni}. The presence of \bps, like certificates, not only signals the existence of \dnscensorship but also pinpoints the actors. However, among all the \dnscensorship detected by \platform, 82.39\% observe invalid certificates without \bps. Therefore, \bp information alone is not enough. Moreover, we discover and publish fingerprints for \bpcnt new \bps that are not covered by previous open-sourced databases. In later sections (\S\ref{sec:res}), we show how \bp fingerprints and properly designed certificate validation in tandem can shed light on actors of \dnscensorship.

\subsection{Comparing with Consistency-based Heuristics}
\label{sec:eva_stat}

As shown in Table~\ref{tab:previous_work}, all state-of-the-art DNS manipulation detection systems incorporate a ``test vs. control'' strategy---comparing potentially censored responses and their metadata with responses from a set of trusted resolvers. Unfortunately, due to an increasingly complex Internet ecosystem, subsequent requests to a single domain even from unmanipulated networks may return different IPs, or different site content, due to complexities in load balancing, CDN deployments, or geo-targeted content serving. To demonstrate how these errors can occur and how our technique compares, we compare against four popular consistency-based techniques: (1) HTTP and Certificate hash matching, (2) AS matching, (3) PTR matching, and finally, (4) statistical thresholding.

When the IP resolved during the test request matches with at least one IP in the control set, we believe that there is no \dnscensorship in place, which is the case in 70.22\% of our DNS resolutions. The aforementioned consistency-based \metrics help cover instances where control IPs fail to account for different points of presence of a given domain. Thus, we find that 28.78\% of test requests did not return an IP in our control set. We apply the above consistency-based \metric comparisons to these 28.78\% of cases using metadata from Censys and Maxmind~\cite{censyspaper,maxmind}.

Overall, we observe that \fnr of true manipulated responses---having an invalid certificate or matching a blockpage fingerprint---are erroneously tagged as correct resolution using consistency-based comparisons \ie 9.70\% of the cases are false negatives. Moreover, a staggering number of \fpr DNS resolutions that are tagged as ``manipulation`` by consistency-based \metrics are false positives.

\myparagraph{Investigating false negatives in consistency-based heuristics}
We provide a breakdown of the false negatives of each of the consistency-based \metrics below (see Table~\ref{tab:iris_fn} for an overview).

%We obtain the HTTP hash and certificate hash information from Censys~\cite{censyspaper}, which stores this information for every IP address on the Internet on common ports

\textit{AS and PTR (CDN) Matching:}
Prior work heavily relies on additional consistency checks based on AS details (name or number) and also performs PTR lookups to check whether the IP address served sits in the same CDN or cloud provider as control IPs. Unfortunately, these metrics are frequently flawed, acting as the major source of false negatives. In our experiments, among all the false negative results, 87.02\% have a matching control AS name or AS name. Almost all of the false negatives have matching control CDN names, as shown in Table~\ref{tab:iris_fn}. This is because some filtering device vendors, like Securly and Infoblox, serve their \bps on IPs in big CDNs (\eg Amazon), which would then be erroneously flagged as \textit{not} \dnscensorship.

\textit{HTTP and Certificate Hash Matching:}
The false negatives introduced by control HTTP hash and certificate hash matching are because previously proposed techniques do not perform sanity checks for the control contents. Conceptually, matching certificate hashes or HTTP hashes (fetched from Censys~\cite{censyspaper}) between the control and test IP addresses can be strong signals of correct DNS resolution. However, some CDNs return a general CDN certificate and an error page when issued a malformed request or a request for an IP address instead of a domain name. 

We observe that our control resolutions sometimes point to these CDNs, and the HTTP and certificate hashes that are stored in these cases are of these error pages and general CDN certificates. These sometimes match with the HTTP and certificate hashes obtained during the test resolution, even though the resolution itself is incorrect. These cases can arise when the manipulated content is hosted on the same network as the legitimate content. We confirm using \platform that these resolutions are incorrect based on sending a query for the \code{(IP, domain)} pairs. Among all the false negatives observed, 3.12\% see an HTTP hash match and 3.85\% see a certificate hash match. 

\begin{table}[t!]
\centering
\small
\refstepcounter{table}

\begin{tabular}{ccccc} 
\toprule
ASN    & AS Owner & Count & Percentage & Type                   \\ 
\midrule
AS3303   & Swisscom     & 86,115 & 13.63\%    & CDN             \\
AS9498   & Airtel       & 82,099 & 13.00\%    & CDN             \\
AS20940  & Akamai       & 63,592 & 10.07\%    & CDN             \\
AS1299   & Arelion      & 33,763 & 5.35\%     & CDN             \\
AS139341 & Aceville Pte & 18,183 & 2.88\%     & Cloud Provider  \\
AS54113  & Fastly       & 16,153 & 2.56\%     & CDN             \\
AS24940  & Hetzner      & 12,524 & 1.98\%     & Cloud Provider  \\
AS9121   & Türk Telekom & 11,815 & 1.87\%     & Telecom  \\
AS9002   & RETN         & 10,380 & 1.64\%     & Telecom     \\
\bottomrule
\end{tabular}
\caption{\textbf{Characterizing false positives of consistency-based \metrics through AS distribution}---The distribution of the top 10 ASes whose IPs are misclassified as manipulation by consistency-based hueristics.}
\label{table:fp_as_distro}
\end{table}

\textit{AS Consistency Thresholding:}
For IPs that are not in the same AS as control IPs, other work has proposed using fine-tuned thresholds~\cite{iclab,gill2015characterizing} to observe how many websites resolve to the same IP. For example, if a set of websites resolve to a single IP from test vantage points, but resolve to IPs in more than $\theta$ ASes from the control nodes, then the test responses for those websites are flagged as \dnscensorship.

To evaluate how effective this thresholding scheme is when compared to checking the certificate validity, we plot the results for DNS resolutions at the global, China, and Indonesia level at each threshold (Figure~\ref{fig:thres_res}). We find that such a threshold method is only capable of identifying a specific kind of \dnscensorship (\eg the \dnscensorship in Indonesia), but omits others (\eg the \dnscensorship in China). Overall, the false negative rate on the global scale is over 50\% and increases as the false positive rate drops. Like other consistency \metrics, we find this thresholding metric to be fragile and introduce errors into \dnscensorship results.

Overall, 9.7\% of the responses marked as legitimate DNS resolutions by the combination of the above consistency-based metrics return an \textit{invalid} certificate or match a \bp as detected by \platform. As shown in Table~\ref{tab:iris_fn}, AS and CDN control matching is the major source of false negatives. Despite the community recognizing AS matching as the most powerful consistency \metric (see Table~\ref{tab:previous_work}), it cannot detect filtering devices hosting their \bps on major CDNs.

\myparagraph{Investigating False Positives in consistency-based heuristics}
To investigates cases where consistency-based \metrics falsely label correct DNS resolutions as manipulated,  we compare our results holistically against all the aforementioned metrics taken in tandem, as this is the approach taken in previous work to detect DNS manipulation~\cite{iris, censoredplanet, ooni}. In particular, we investigate how the collective determination of consistency-based \metrics (i.e., manipulated or unmanipulated) differ from \platform's determinations.

As shown in Table~\ref{tab:anomaly_comp}, we observe that the \fpr of the DNS responses tagged as manipulated (\ie do not match with \textit{any} control data) contain IP addresses that host a valid certificate for the queried domains, which we consider as false positives. 

In investigating these false positives, we uncover that the vast majority of IPs that were tagged as \dnscensorship are hosted by CDNs and ISPs (Table~\ref{table:fp_as_distro}). However, these CDNs and ISPs are not known \censors---rather, they may deploy highly distributed resolvers that return IPs based on a number of different decisions (\eg anycast, load balancing) which the consistency metrics proposed by previous work do not adequately capture.

In addition, there are two other reasons for the presence of these false positives:
\begin{enumerate}
    \item The coverage of control resolvers is always limited. Although we report 70.22\% effectiveness of using matching control IPs to identify unmanipulated responses, the control groups fail to cover all the points of presence.
    \item All metadata information used for consistency checks (e.g., AS information, HTTP hashes, and certificate validation) are collected from auxiliary databases (Censys and Maxmind) in our work as well as previous work~\cite{iris, censoredplanet} and may be incomplete. For the IPs returned by test resolvers in our measurements, only 30.3\% have a certificate hash, 93\% have an HTTP hash, and 99\% have a matching AS name and AS number.
\end{enumerate}

% \begin{table}[t]
%     %\scriptsize
%     \centering
% \begin{tabular}{ | m{10em} | m{14em}| } 
%         \hline
%         {OpenDNS IP} & {Hostname}  \\
%         \hline
%          146.112.61.105 & hit-botnet.opendns.com\\
%         \hline
%         146.112.61.106 & hit-adult.opendns.com  \\
%         \hline
%         146.112.61.107 & hit-malware.opendns.com  \\
%         \hline
%         146.112.61.108 & hit-phish.opendns.com\\
%         \hline
%         {others} & hit-block.opendns.com\\
%         \hline
%     \end{tabular}

% \end{table}

\begin{table}[t!]
\centering
\small
\vspace{-5 pt}
\begin{tabular}{m{13em} | m{14em}}
\toprule
{OpenDNS IP} & {Hostname} \\ 
\midrule
146.112.61.105 & hit-botnet.opendns.com\\
\midrule
146.112.61.106 & hit-adult.opendns.com  \\
\midrule
146.112.61.107 & hit-malware.opendns.com  \\
\midrule
146.112.61.108 & hit-phish.opendns.com\\
\midrule
{others} & hit-block.opendns.com\\
\bottomrule
\end{tabular}
\caption{IPs owned by OpenDNS detected by \platform.}
\label{tab:cisco}
\end{table}

In summary, by comparing \platform's findings across prior work's consistency metrics, we demonstrate how identifying \dnscensorship via IP metadata matching can provide fragile and sometimes incorrect results---\fpr of the manipulated detected by the consistency metrics are false positive, and \fnr manipulated DNS responses are omitted. In contrast, certificate validation and \bp matching provide a clear improvement in accuracy. 

\section{Findings}
\label{sec:res}

\label{sec:res_isp}

%\platform identifies about 200 unique \bp fingerprints from clustering the HTTP(S) pages hosted on the IPs returned by global open resolvers. By leveraging certificate validation, we deepen our understanding of the global deployment of \dnscensorship. 

In this section, we describe the key findings we observe from investigating \dnscensorship with certificate validation. \platform discovers commercial filtering product deployment in \tlsproxycountrycnt countries, as well as ISP-level \dnscensorship in \ispcountrycnt countries, with a wide diversity of \dnscensorship deployment strategies. 82.39\% of the invalid certificates we detect come without a \bp. To the best of our knowledge, this is the first report of the implementers of \dnscensorship on a global scale. Through the lens of certificate validation and \bp matching, we are not only able to detect signals of \dnscensorship but also pinpoint the actors.

\begin{table}[t]
    \centering
    \small

\begin{tabular}{m{0.1cm} m{1.3cm} m{0.5cm} m{0.5cm} m{0.5cm} m{3cm}} 
        \toprule
        {\cellcolor{white}} & {\cellcolor{white}Product} & {\cellcolor{white}Origin} & {\cellcolor{white}Block Page} & {\cellcolor{white}Root Cert} & {\cellcolor{white}Country of Deployment} \\
        \midrule
        \multirow{9}{0em}{\rotatebox[origin=c]{90}{Observed in one country}} 
        & Canadian Shield 
        & CA 
        & \mycircle{pastelred}
        & \mytriangle{calpolypomonagreen}
        & CA \\
        
        & \cellcolor{lightcyan} WebTitan     
        & \cellcolor{lightcyan} US  
        & \cellcolor{lightcyan} \mycircle{pastelred}
        & \cellcolor{lightcyan} \mytriangle{pastelred}
        & \cellcolor{lightcyan} US \\
        
        & OneDNS 
        & CN 
        & \mycircle{pastelred}
        & \mytriangle{calpolypomonagreen}
        & CN \\
        
        & \cellcolor{lightcyan} JusprogDNS 
        & \cellcolor{lightcyan} DE 
        & \cellcolor{lightcyan} \mycircle{pastelred}
        & \cellcolor{lightcyan} \mytriangle{pastelred}
        & \cellcolor{lightcyan}DE \\
        
        & Infoblox 
        & US 
        & \mycircle{pastelred}
        & \mytriangle{pastelred}
        & US\\
        
        & \cellcolor{lightcyan} NextDNS 
        & \cellcolor{lightcyan} US 
        & \cellcolor{lightcyan}	\mycircle{pastelred}
        & \cellcolor{lightcyan} \mytriangle{pastelred}
        & \cellcolor{lightcyan}US  \\
        
        & Comodo 
        & US 
        & \mycircle{pastelred}
        & \mytriangle{calpolypomonagreen}
        & US\\
        
        & \cellcolor{lightcyan} Zyxel 
        & \cellcolor{lightcyan} CH 
        & \cellcolor{lightcyan} \mycircle{pastelred}
        & \cellcolor{lightcyan} \mytriangle{pastelred}
        & \cellcolor{lightcyan} CH\\
        
        & WatchGuard 
        & US 
        & 	
        & \mytriangle{pastelred}
        & US\\
        
        & \cellcolor{lightcyan} Securly 
        & \cellcolor{lightcyan} US 
        & \cellcolor{lightcyan} \mycircle{pastelred}
        & \cellcolor{lightcyan} \mytriangle{pastelred}
        & \cellcolor{lightcyan} US\\
        \midrule

        \multirow{7}{14em}{\rotatebox[origin=c]{90}{Observed in multiple countries}} 
        &OpenDNS (Cisco)     
        & US 
        & \mycircle{pastelred}
        & \mytriangle{pastelred}
        & AR, AU, BR, CA, CL, CN, CR, CZ, DE, ES, FR, GR, ID, IE, IN, IT, JP, KR, KZ MX,  NZ,  RO, SE, SK, US, ZA  \\
        
        & \cellcolor{lightcyan}AdguardDNS
        & \cellcolor{lightcyan} CA 
        & \cellcolor{lightcyan} \mycircle{pastelred}
        & \cellcolor{lightcyan} \mytriangle{calpolypomonagreen}
        & \cellcolor{lightcyan} GB, BY, CY, FR, ID, LV, NZ, RU\\
        
        & SafeDNS 
        & US 
        & \mysquare{pastelred}
        & \mytriangle{pastelred}
        & AU, NL, US \\
        
        & \cellcolor{lightcyan} Kewoen 
        & \cellcolor{lightcyan} DE 
        & \cellcolor{lightcyan} 
        & \cellcolor{lightcyan} \mytriangle{pastelred}
        & \cellcolor{lightcyan} AU, DE, FR, GB, JP, NL, US \\
        
        & SkyDNS 
        & RU 
        & \mysquare{pastelred}
        & \mytriangle{pastelred}
        & RU, UA, KZ\\
        
        & \cellcolor{lightcyan} CloudVeil 
        & \cellcolor{lightcyan} US 
        & \cellcolor{lightcyan} \mycircle{pastelred}
        & \cellcolor{lightcyan} \mytriangle{pastelred}
        & \cellcolor{lightcyan}CA, US\\
        
        & Fortinet 
        & US 
        & 
        & \mytriangle{calpolypomonagreen}
        & AR, AT, AU, BD, BF, BR, CA, CH, CL, CN, CZ, DE, DK, FR, GB, HK, ID, IN, IQ, IT, JP, KR, KW, MR, MY, NL, PH, PL, SV, TH, TR, TT, TW, US\\
        \bottomrule
    \end{tabular}
        \caption{\textbf{Location of origin and deployment of different filtering products identified via certificate validation and \bp matching} - (1) For the Blockpage column, a red square indicates a legal blockpage, a red circle indicates a blockpage that does not contain legal information. (2) For the Root Cert column, a green triangle indicates a trusted root CA, a red triangle indicates an untrusted root CA. }
    \label{tab:prod_res}
\end{table}

\subsection{Identifying Filtering Product Vendors}
%\elisa{identify the actors themselves, instead of merely knowing it is \dnscensorship, we know the devices that conduct TLS interception}

\label{sec:res_prod}
\platform identifies \tlsproxycnt \dnscensorship filtering product vendors deployed in \tlsproxycountrycnt countries, as shown in Table~\ref{tab:prod_res}. Most (94.11\%) commercial filtering devices return an IP hosting (configurable) \bps. Vendors deploy different strategies for \dnscensorship. For example, certificate chains returned by IPs owned by Cira, OneDNS, AdguardDNS, and Fortinet have a trusted root CA \eg DigiCert and ZeroSSL. In this case, the common name of the certificates is usually issued for the product website (\eg \code{*.onedns.net}). Other products attempt to perform a man-in-the-middle \ie the leaf certificate has a matching hostname with the queried domain, yet the root CA is not trusted by major browsers.

%To the best of our knowledge, this is the first report of products performing \dnscensorship on the global scale. 

We discover 10 vendors whose products are only seen in the country of origin. We also observe a concerning pattern where \dnscensorship vendors export their products to countries without such technology, similar to findings from previous work on the spread of HTTP filtering devices \cite{filtermap,dalek2013method}. SkyDNS, a vendor based in Russia, claims that their technology is already used in more than 45 countries, advertised as ``Solutions for country wide content filtering and network security''~\cite{skydns}. \platform captures \bps with 451~status code (``Unavailable For Legal Reasons'') from SkyDNS products and SafeDNS, implying that they are targeting ISPs and governments as potential consumers.

We observe that 7 commercial filtering products with multinational coverage return a common set of IP addresses in all countries, indicating that these IP addresses are managed centrally. For example, among all the DNS responses from  resolvers in 34 countries that are tampered by Fortinet, only one IP is observed (\code{208.91.112.55}), which sits in AS40934 (Fortinet Inc).  Interestingly, OpenDNS, Cisco's DNS subsidiary, owns a small pool of anycast IPs whose hostname indicates the exact content they are blocking (Table~\ref{tab:cisco}). This observation can potentially be used in \dnscensorship circumvention. If the middleboxes are \textit{injecting} the tampered IPs, the DNS client can be modified to discard a known pool of \dnscensorship product IPs to wait for the correct resolution to arrive.

\subsection{Identifying ISP DNS Manipulation}

\begin{figure} 
 \centering
 \includegraphics[width=0.99\columnwidth]{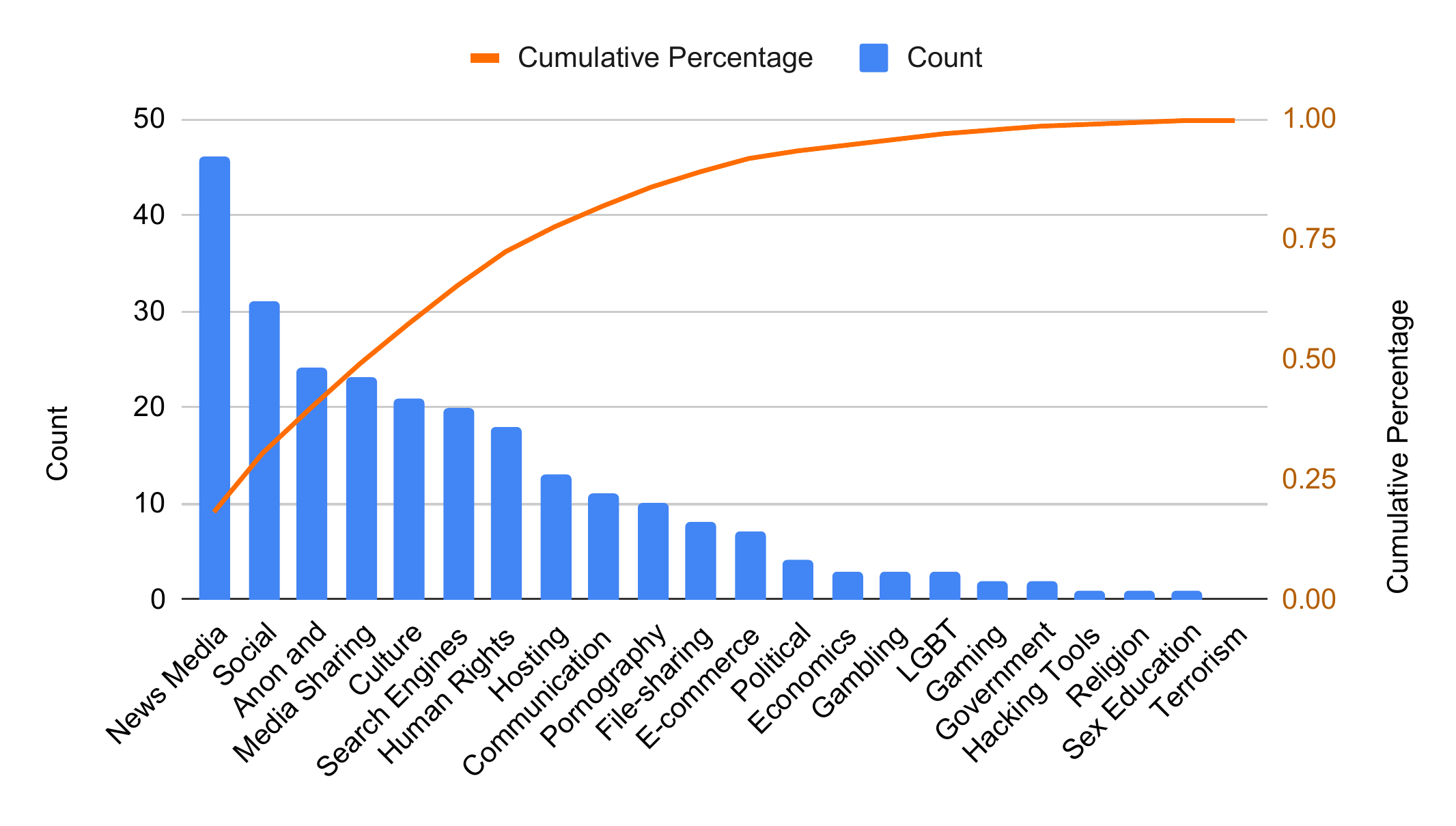}
 \caption{\textbf{Category Distribution of domains blocked in China} - \platform detects \dnscensorship in China via certificate validation.}
\label{fig:cb_blocked_domains}
\end{figure}

\begin{figure}
\hfill
\includegraphics[width=4cm]{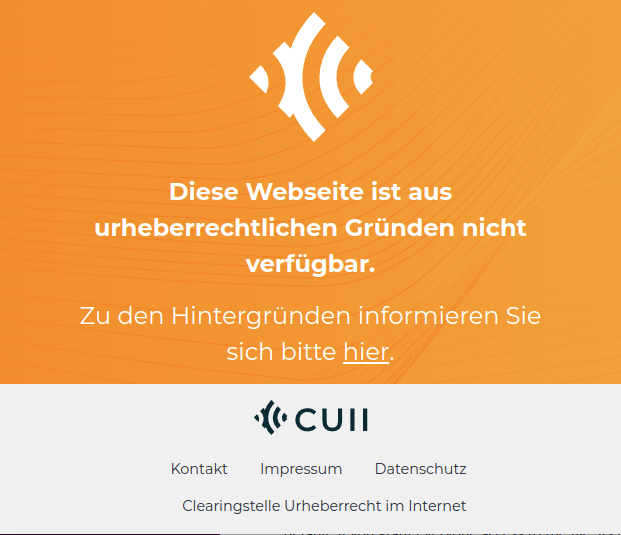}
\hfill
\includegraphics[width=4cm]{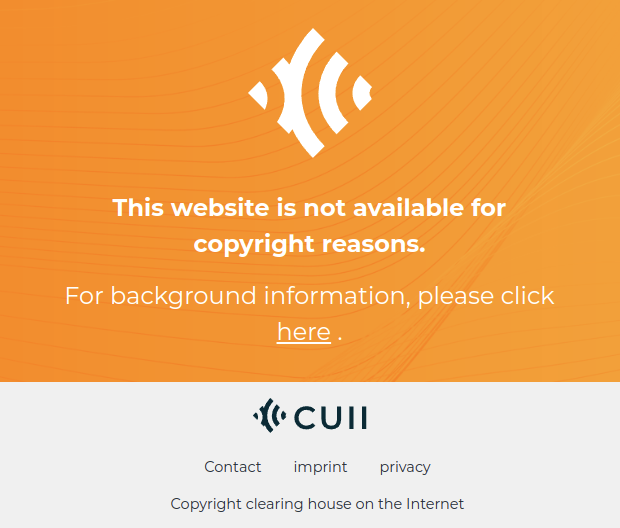}
\hfill
\caption{\textbf{Copyright infringement \bp from CUII}---a joint initiative of affected German industry associations and ISPs \protect\cite{bundesnetzagentur}}
\label{fig:cuii}
\end{figure}

\platform detects \dnscensorship on the ISP level in \ispcountrycnt countries via certificate validation, ranging from previously well-studied countries in Internet censorship \eg Russia \cite{decentralized, ognyanova2019putin}, to countries that previous research in Internet censorship did not investigate in depth, \eg Indonesia, Nepal, Thailand, and Romania. We also see ISPs performing \dnscensorship in countries that Freedom House classified as ``Free'' \cite{freedomhouse}, such as Germany, Greece, and Denmark. ISPs in these countries usually return a \bp indicating blocking of copyright-infringing domains (as shown in Figure~\ref{fig:cuii}). In ``Not Free'' countries such as China, Russian, and Iran, the blocked categories include news media, circumvention tools, social media, and other major services \eg Google and Wikipedia (as shown in Figure~\ref{fig:cb_blocked_domains}). \platform's dataset also covers signals of \dnscensorship implemented by companies, universities, and other organizations.

We observe heterogeneity in ISP-level \dnscensorship even for ISPs within the same country. As shown in Table~\ref{tab:isp_res}, ISPs can either use services like Let's Encrypt, DigiCert, or Sectigo to issue certificates trusted by major browsers for their \bps, or simply issue self-signed certificates, which are easy to create and do not involve any financial cost. Most ISPs return a \bp for blocked contents, either citing the laws and regulations that legitimize such blocking or simply state the access is forbidden or denied in their local languages. However, we observe that some of them just return a blank page or a default setup page of OpenResty. The common name and issuer field in the certificates issued by ISPs are informative of \dnscensorship as well. Some indicate the ISPs are the issuers of the certificates, others even indicate that the certificates are issued by the ISP for the purpose of \dnscensorship.

\begin{table}[t]
    \small
    \centering
    \renewcommand{\arraystretch}{1.2}
    
    \aboverulesep = 0mm
    \belowrulesep = 0mm
\begin{tabular}{ m{1.2cm}  m{3.7cm} m{0.5cm} m{0.5cm} m{0.5cm} }
    \toprule
    {\cellcolor{white}Country}  &  {\cellcolor{white}AS number of returned IPs} & {\cellcolor{white} Leaf Cert} & {\cellcolor{white}Block Page} & {\cellcolor{white}Root Cert} \\

        \midrule
        \multirow{6}{0em}{Russia} 
        & \cellcolor{lightcyan}AS12616, AS44347, AS44587,\newline AS49505, AS34241 
        & \cellcolor{lightcyan} \mysquare{pastelred} 
        & \cellcolor{lightcyan} \mysquare{pastelred}	 
        & \cellcolor{lightcyan} \mytriangle{calpolypomonagreen} \\
        
        & AS25549, AS31483, AS34757 
        & \mycircle{pastelred} 
        & \mycircle{pastelred}
        & \mytriangle{calpolypomonagreen} \\
        
        &\cellcolor{lightcyan} AS12389, AS50466
        &\cellcolor{lightcyan} 
        &\cellcolor{lightcyan} \mysquare{pastelred} 
        &\cellcolor{lightcyan} \mytriangle{pastelred}\\
        
        & AS42071
        & \mycircle{pastelred}
        &&\mytriangle{pastelred}\\
        
        &\cellcolor{lightcyan} AS57571, AS43287, AS49469
        &\cellcolor{lightcyan} \mysquare{pastelred} 
        &\cellcolor{lightcyan} \mysquare{pastelred} 
        &\cellcolor{lightcyan} \mytriangle{pastelred}\\
        
        & AS8395
        &\mycircle{pastelred} 
        &\mysquare{pastelred} 
        &\mytriangle{pastelred}\\
        \midrule
        
        \multirow{2}{0em}{Ukraine} 
        & \cellcolor{lightcyan} AS42546
        & \cellcolor{lightcyan} \mycircle{pastelred} 
        & \cellcolor{lightcyan} \mysquare{pastelred}	 
        & \cellcolor{lightcyan} \mytriangle{calpolypomonagreen} \\
        
        & AS42546
        & \mycircle{pastelred} 
        & \mycircle{pastelred}	 
        & \mytriangle{calpolypomonagreen} \\

        \midrule
        \multirow{6}{0em}{Indonesia} 
        & \cellcolor{lightcyan} AS58396, AS45287, AS38758
        & \cellcolor{lightcyan} \mycircle{pastelred} 
        & \cellcolor{lightcyan} \mycircle{pastelred}	 
        & \cellcolor{lightcyan} \mytriangle{pastelred} \\
        
        & AS9341, AS5578
        &
        & \mycircle{pastelred} 
        & \mytriangle{pastelred}\\

        & \cellcolor{lightcyan} AS16276, AS141626, AS7713
        & \cellcolor{lightcyan} \mycircle{pastelred} 
        & \cellcolor{lightcyan} \mycircle{pastelred} 
        & \cellcolor{lightcyan} \mytriangle{pastelred}\\
        
        & AS58495, AS132634
        & \mysquare{pastelred}
        &
        & \mytriangle{calpolypomonagreen}\\
        
        & AS140413, AS136873
        & \mysquare{pastelred}
        & \mysquare{pastelred}
        & \mytriangle{calpolypomonagreen}\\

        & \cellcolor{lightcyan} AS56241
        & \cellcolor{lightcyan} \mycircle{pastelred} 
        & \cellcolor{lightcyan} \mycircle{pastelred} 
        & \cellcolor{lightcyan} \mytriangle{calpolypomonagreen}\\
        
        \midrule
        
        \multirow{2}{0em}{Nepal} 
        & AS63991
        & \mycircle{pastelred} 
        & 
        & \mytriangle{calpolypomonagreen} \\
        
        & \cellcolor{lightcyan} AS140973
        & \cellcolor{lightcyan} \mysquare{pastelred} 
        & \cellcolor{lightcyan} \mycircle{pastelred}	 
        & \cellcolor{lightcyan} \mytriangle{calpolypomonagreen} \\

        \midrule
        \multirow{1}{0em}{Thailand} 
        & AS23969
        & \mysquare{pastelred} 
        &
        & \mytriangle{calpolypomonagreen} \\
        
        \midrule
        \multirow{1}{0em}{Singapore} 
        & \cellcolor{lightcyan} AS3758, AS3758
        & \cellcolor{lightcyan} \mycircle{pastelred} 
        & \cellcolor{lightcyan} 
        & \cellcolor{lightcyan} \mytriangle{pastelred} \\
        
        \midrule
        \multirow{1}{0em}{Belarus} 
        & AS6697
        & 
        & \mysquare{pastelred} 
        & \mytriangle{pastelred} \\
        
        \midrule
        \multirow{1}{0em}{Lithuania} 
        & \cellcolor{lightcyan} AS212531
        & \cellcolor{lightcyan} \mycircle{pastelred} 
        & \cellcolor{lightcyan} \mysquare{pastelred} 
        & \cellcolor{lightcyan} \mytriangle{calpolypomonagreen} \\
        
        \midrule
        \multirow{2}{0em}{Romania} 
        & AS31313
        & 
        & \mysquare{pastelred} 
        & \mytriangle{pastelred} \\

        & \cellcolor{lightcyan} AS12302
        & \cellcolor{lightcyan} 
        & \cellcolor{lightcyan} 
        & \cellcolor{lightcyan} \mytriangle{pastelred} \\
        \midrule

        \multirow{2}{0em}{Belgium} 
        & AS2611
        & \mycircle{pastelred} 
        & \mycircle{pastelred} 
        & \mytriangle{calpolypomonagreen} \\

        & \cellcolor{lightcyan} AS5432, AS8717
        & \cellcolor{lightcyan} \mycircle{pastelred} 
        & \cellcolor{lightcyan} 
        & \cellcolor{lightcyan} \mytriangle{pastelred} \\
        \midrule
        \multirow{1}{0em}{Denmark} 
        & AS35158
        & \mycircle{pastelred} 
        & \mycircle{pastelred} 
        & \mytriangle{calpolypomonagreen} \\
        \midrule
        
        \multirow{1}{0em}{Italy} 
        & \cellcolor{lightcyan} AS29050
        & \cellcolor{lightcyan} \mysquare{pastelred} 
        & \cellcolor{lightcyan} 
        & \cellcolor{lightcyan} \mytriangle{pastelred} \\
        
        %\midrule
        %\multirow{1}{0em}{Columbia} 
        %& AS35158
        %& \mycircle{pastelred} 
        %& \mysquare{pastelred} 
        %& \mytriangle{calpolypomonagreen} \\
        %\midrule
        
        \multirow{1}{0em}{Greece} 
        & \cellcolor{lightcyan} AS6799
        & \cellcolor{lightcyan} \mysquare{pastelred} 
        & \cellcolor{lightcyan} 
        & \cellcolor{lightcyan} \mytriangle{pastelred} \\
        
        \midrule
        \multirow{1}{0em}{Switzerland} 
        & AS3303
        & \mycircle{pastelred} 
        & \mysquare{pastelred} 
        & \mytriangle{calpolypomonagreen} \\
        
        \midrule
        \multirow{1}{0em}{Germany} 
        & \cellcolor{lightcyan} AS24940
        & \cellcolor{lightcyan} 
        & \cellcolor{lightcyan} \mysquare{pastelred} 
        & \cellcolor{lightcyan} \mytriangle{pastelred} \\
        
        \midrule
        \multirow{1}{0em}{Australia} 
        & AS16509
        & \mysquare{pastelred} 
        & \mysquare{pastelred} 
        & \mytriangle{calpolypomonagreen} \\
        
        \bottomrule
    \end{tabular}
    \caption{\textbf{Countries where \platform detects ISP-level \dnscensorship via certificate validation}---(1) For the Leaf Cert column, a red square indicates that the ISP is specified as the issuer and the certificate is issued for blocking. A red circle indicates only that the certificate is issued by the ISP. (2) For the Blockpage column, a red square indicates a legal blockpage, and a red circle indicates a blockpage that does not contain legal information. (3) For the Root Cert column, a green triangle indicates a trusted root CA, and a red triangle indicates an untrusted root CA.}
    \label{tab:isp_res}
\end{table}

For example, in Russia, we see tampered IPs in 16 different ASes returned by ISP-owned resolvers. \platform identifies 31 unique \bp fingerprints in Russia, revealing the decentralized nature of Russia's national \dnscensorship~\cite{decentralized}. We see the presence of badly configured certificates that have no information in all the fields except the effective date, expiration date, and the country of the issuer (``\code{RU}``). We also observe carefully configured certificates that specify both the ISP and the purpose of the certificate in the common name \eg \code{forbidden.citytelecom.ru}, issued by an ISP based in Moscow. \dnscensorship by ISPs can be quite obscure if the implementer chooses to not return a \bp and does not configure an informative certificate. For example, in Romania, we see a certificate with an IP address in the common name (\code{213.177.28.90}) returned with the HTTPS page. The IP address in the certificate common name hosts a \bp stating that the access to the requested website is blocked by the decision of the Supervisory Committee of the O.N.J.N, the gambling regulation institution of Romania. Some ISPs only return a default webpage like ``Welcome to nginx!`` along with their ISP certificate. Therefore, checking \bp matching and certificates in tandem helps us to have a more holistic view of overt \dnscensorship.

We also discover ISP-level \dnscensorship via HTTP-only \bp matching in countries including Russia, Indonesia, Turkey, Poland, Italy, Romania, India, Columbia, Belgium, Philippines, Mexico, Australia, Nepal, and Ukraine. Examples of ISP-level HTTP \bps without certificates are shown in Fig.~\ref{fig:http_only_isp_bp} in Appendix \ref{sec:app_http_only}. In total, \platform discovers ISP-level \dnscensorship in \ispcountrycnt countries.

\subsection{Identify Covert \dnscensorship}
\label{sec:res_covert}

% \begin{table}[t]
%     %\scriptsize
%     \centering
% \begin{tabular}{ | m{6em} | m{3em}|m{3em}|m{3.5cm}| } 
%         \hline
%         {Country/Region} & {Hong Kong} & {Singa-pore} & {Bangladesh,  South Korea, Indonesia, Myanmar, Thailand}  \\
%         \hline
%          Overlap & 85.43\% & 85.96\% & 100\%\\
%         \hline
%          Domains & 198 & 174 & < 5 \\
%         \hline
%     \end{tabular}
%     \caption{Countries/Regions that are potentially affected by China's censorship leakage.}
%     \label{tab:cn_collateral}
% \end{table}

\begin{table}[t]
    \small
    \centering
    
\begin{tabular}{m{6em} m{3em} m{3em} m{3.5cm}} 
        \toprule
        {Country/Region} & {Hong Kong} & {Singa-pore} & {Bangladesh,  South Korea, Indonesia, Myanmar, Thailand}  \\
        \midrule
         Overlap & 85.43\% & 85.96\% & 100\%\\
        \midrule
         Domains & 198 & 174 & < 5 \\
        \bottomrule
    \end{tabular}
    \caption{\textbf{Countries/Regions that are potentially affected by China's censorship leakage}---The overlap indicates the fraction of \code{(domain, resolved IP)} returned by affected resolvers outside China that overlap with Chinese \dnscensorship.}
    \label{tab:cn_collateral}
\end{table}

%\elisa{do we want to talk about how China has a fail-safe for those US IPs?}
From analyzing the heterogeneity of \dnscensorship practice by commercial products and ISP deployment, we learn that it is important to integrate both the information inferred from \bps and certificates. In the worst case, a \censor can choose to issue a non-informative certificate with no \bps, making it very hard to determine the implementer and purpose of \dnscensorship, giving the \censor itself plausible deniability of implementing Internet blocking. In this subsection, we will investigate a case of covert \dnscensorship discovered by certificate validation.

Among all the invalid certificates \platform detected, 82.39\% come without a \bp. %About 60.82\%/. 
About 39.17\% of the invalid certificates do not contain information about filtering product vendors or ISPs, making it more covert. The certificates in this category have trusted root CAs and mismatching hostnames, coming with an HTTPS page with a client error status code (400+), indicating that the queried HTTPS servers either can not find the resource the user requested, or think it is a bad request.  98.66\% of those IPs are returned by DNS resolvers in China. The returned IPs belong to a huge IP pool located in ASes owned by Facebook, Twitter, Cloudflare, and other blocked CDN providers, which confirms the finding in previous research \cite{chinatriplet, marczak2015analysis, greatgfw, hoang2019measuring, wang2017your} that China is injecting IP addresses belonging to popular US companies. The categories of domains blocked in China are shown in Fig.~\ref{fig:cb_blocked_domains}. Resolvers in 14 other countries and regions share this behavior, seven of which contain at least one resolver that sees a complete overlap of the same \dnscensorship methodology as the Chinese GFW, as shown in Table~\ref{tab:cn_collateral}. China's DNS injections are sometimes cached by resolvers outside China, despite the administrators of those DNS resolvers having no intention to implement \dnscensorship \cite{hoang2019measuring, nebuchadnezzar2012collateral, Nosyk23a}.

0.6\% of the certificates that come with a status code 400+ are issued by Cira, a DNS firewall ``to block access to malicious websites`` \cite{cira}, shown in Table~\ref{tab:prod_res}. As another case of covert DNS manipulation, we observe countries such as Iran utilizing private IPs to block sensitive content. This form of \dnscensorship is more opaque than the previously discussed cases. It is very hard to determine whether the \dnscensorship is intentional when no \bps are served and the traffic is diverged to the IPs that are not owned by the \censors. Judging if the blocking is intentional is a hard task in covert \dnscensorship,  but by checking the validity of certificates returned, we are able to obtain the users' perspective and understand if the right resource is hosted on the returned IPs.

\subsection{Case Study: IPs With No Certificates}
\label{sec:conerr}

Investigating cases where test revolvers fail for both HTTP and HTTPS requests produces indicators of misconfigured resolvers (i.e. all domains tested resolved to the same IP and subsequently fail for both HTTP and HTTPS) but also indicators of revolvers configured for specific domain blocking. For example, we discover 83 Russian resolvers that assign between 20 and 114 domains to the IP \code{62.33.207.197}. The domains assigned to this IP by the resolver include \code{bbc.com}, \code{bridges.torproject.org}, and \code{psiphon.ca}. Upon investigation, we discover that port 80 and port 443 on this IP address are closed; the only open port is port 444, which returns a Russian blockpage (Figure~\ref{fig:ttk}). \platform does not aim to scan all the potential open ports of the returned IPs. However, by filtering out resolvers that return the same IP for multiple queried domains, we obtain potential signals of \dnscensorship that can be confirmed by further analysis.

\section{Limitations}
\label{sec:discuss}

In this paper, we only consider \dnscensorship on the conventional port 53 (as well as page fetch from port 80 for HTTP and port 443 for HTTPS). However, previous work has shown that \dnscensorship can happen on other ports as well~\cite{pearcemany}. Moreover, we do not consider the rare possibility of rogue certificate authorities \cite{tung2016mozilla, prins2011diginotar, raman2020investigating}. It is possible for nation-state actors to conduct such silent MitM attacks while evading our detection. We rely on major browsers such as Mozilla Firefox and Google Chrome to remove such root CAs from their trust lists.

Censors could employ unreachable IPs to prevent users from accessing the domains they request, using either (1) private IPs or (2) public IPs that do not host anything on port 80 or port 443. We identify \dnscensorship performed using private IPs, but choose to mark the second case as \textit{unknown}. Future work can manually investigate those IPs to identify whether these cases are \dnscensorship. A case study of such an investigation can be found in Appendix~\ref{sec:app_connect_error}. Moreover, for the case where the IPs only host an HTTP page, we do not attempt to identify real pages by comparing page content. Instead, we match HTTP pages with \bp fingerprints. Web services often have country-specific dynamic content which could lead to inaccuracies.

While this study includes 6 months of data from Censored Planet, it is worth noting that further longitudinal analysis of DNS data has the potential to capture useful signals regarding emerging patterns and trends in DNS manipulation. Diversifying the geolocation of vantage points for DNS resolution measurement and HTTP(S) page fetch can yield data that more accurately reflect the user experience of DNS censorship. Such an approach can provide a more comprehensive understanding of the global patterns and impact of DNS manipulation. We hope future research will analyze the discrepancies between measurements obtained from different vantage points. Such analysis can enable researchers to gain a more complete picture of global content delivery and to identify instances of server-side blocking.
\section{Conclusion}
%This paper presents \platform, a remote measurement platform aiming to detect global \dnscensorship. 
Developing \metrics to accurately detect \dnscensorship on a global scale is challenging. State-of-the-art \metrics introduced by previous work that identify shared infrastructure---though intuitive---are error-prone given the advancements in Internet infrastructure such as CDNs. We discover that \fpr of the manipulated DNS responses identified by the current state-of-the-art are false positives. By taking one step forward to fetch the HTTP(S) pages hosted on IPs returned by resolvers, \platform simulates the users' perspective to understand the accessibility of requested resources. By leveraging certificate validation and blockpage matching, \platform identifies \tlsproxycnt TLS proxy vendors deployed in \tlsproxycountrycnt countries, as well as \ispcountrycnt countries with ISP-level \dnscensorship. From \platform's dataset, we construct \bpcnt unique \bp fingerprints for previously unknown \bps. Our techniques and the curated \bp fingerprints are open-sourced. We have collaborated with Censored Planet~\cite{censoredplanet}, an open censorship measurement platform, to integrate our techniques into its functioning, and we are actively working on integrating our techniques into other measurement platforms such as OONI~\cite{ooni}. We hope our techniques enable accurate detection of \dnscensorship and improve the quality of open-access data provided to the Internet freedom community.

\section{Acknowledgements}
The authors would like to thank the anonymous reviewers for their helpful comments on the paper. We also thank Will Scott, David Wang, and Daniel Liu for their valuable feedback on DNS measurements. This work was supported by the Defense Advanced Research Projects Agency under Agreement No. HR00112190127 and a Bureau of Democracy, Human Rights and Labor (DRL) Grant (No. SLMAQM20GR2132). 

\bibliographystyle{plain}
\bibliography{sample-base}

\appendix
\label{sec:appendix}
\section{Appendix}

\begin{figure}[!t]
 \centering
    \includegraphics[scale=0.35]{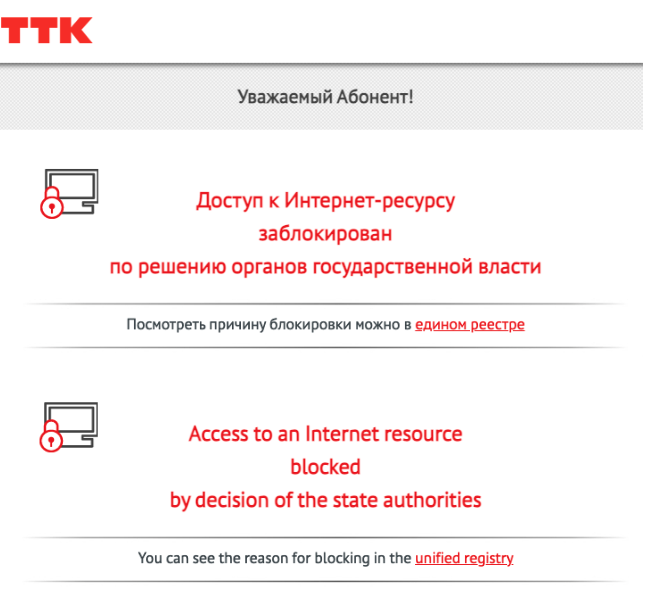}
    \caption{Russian ISP \bp hosted on IP \code{62.33.207.197}, port 444.}
    \label{fig:ttk}
\end{figure}

\subsection{Case Study: Nonzero \rcode}

\label{sec:rcode_case_study}
By properly filtering out erroneous resolvers, we observe both commercial filtering products and ISPs deploying DNS manipulation by using \code{\rcode:3 (NXDOMAIN)}. For example, 4 resolvers (0.*.dns.gamban.com) all return \code{\rcode:3} for exactly 47 domains. The TLD of these resolvers, Gamban, is a commercial filtering product that offers \dnscensorship as a service, which evidently is implemented by returning DNS \code{NXDOMAIN} ~\cite{gamban}.  All but 4 of Gamban's 47 blocked sites are gambling domains. The 4 outliers were all circumvention tools:  \code{tunnelbear.com}, \code{www.ipredator.se},  \code{torproject.org}, and \code{bridges.torproject.org}.  Although we see the locations of the resolvers to be Great Britain, United States, and Singapore, those appear to be the locations of the Gamban servers, which could be requested by users globally. Previous work \cite{iris} found that \code{NXDOMAIN} is relatively infrequently used for \dnscensorship. However, the existence of such middlebox vendors demonstrates the diversity and evolution of \dnscensorship deployment.

We also see several examples of ISP \dnscensorship utilizing \code{\rcode:3, NXDOMAIN}. In one example, we see 65 different ROSTELECOM resolvers all of which return \code{NXDOMAIN} for exactly two domains: \code{www.facebook.com} and \code{staticxx.facebook.com}. In Brazil, six resolvers from Telefonica with the TLD \code{gvt.net.br} return \code{\rcode:3} for 7 sites:\code{piratebay.org}, \code{womenonwaves.org}, and 5 adult content domains. In an example of organizational blocking, we observe 3 resolvers owned by Thai Cyber University using \code{\rcode:3} to block 6 sites, including \code{americannaziparty.com} and \code{nostraightnews.com}. By investigating resolvers that return \code{NXDOMAIN}, meaningful signals of DNS manipulation emerge from our dataset. We discover the use of \code{NXDOMAIN} to deploy DNS manipulation, by a diverse set of actors: commercial vendors, ISPs, and organizations like universities and banks. 

%Among the control pages, we see 92.69\% \elisa{should calculate average among all the snapshots} of 

%\deepak{These are fine limitations, but I think you want to add them after you present the entirety of the system + section, rather than up front.}

\subsection{Annotating DNS responses}

More often than not, DNS resolvers return more than one IP for the queried domain. In these cases, a client will need to decide which of the returned IPs to connect to first. The behavior of the DNS client depends on its implementation. Generally, it tries the IPs in the order they were returned by the DNS server in a round robin manner~\cite{rrclient}. 

%Different implementations of DNS servers have different strategies to choose which IP(s) to return. The ordering of the returned IPs may imply history information the resolver has to the each IP(s) in list, which usually consist of some sort of weighted averages for the response time of the address, and the batting average of the address\cite{rfc1035}. 

In most cases of \dnscensorship, we observe that no legitimate IP for the queried domain is included in the DNS responses. In a few very rare cases ($4e^{-5}\%$), we see mixed signals, where IPs hosting a \bp as well as IPs hosting legitimate content are returned in the same response. This is potentially a case of the collateral damage of DNS poisoning: manipulated DNS records are cached by open resolvers with no intention to block. We mark these cases as unmanipulated in our study.

%Thus, HTTP and HTTPS failure can be helpful in determining both misconfigured resolvers that give false-positives for censorship tallies, but also, resolvers failing for only a curated subset of domains tested. 

\label{sec:app_connect_error}
\subsection{HTTP-Only ISP Blockpages}
\label{sec:app_http_only}

We discover multiple countries with ISP-level \dnscensorship where only HTTP \bps are returned. Below are 6 examples of such \bps with corresponding English translation. 
\begin{figure*}
  \centering

    \includegraphics[width=1.9\columnwidth]{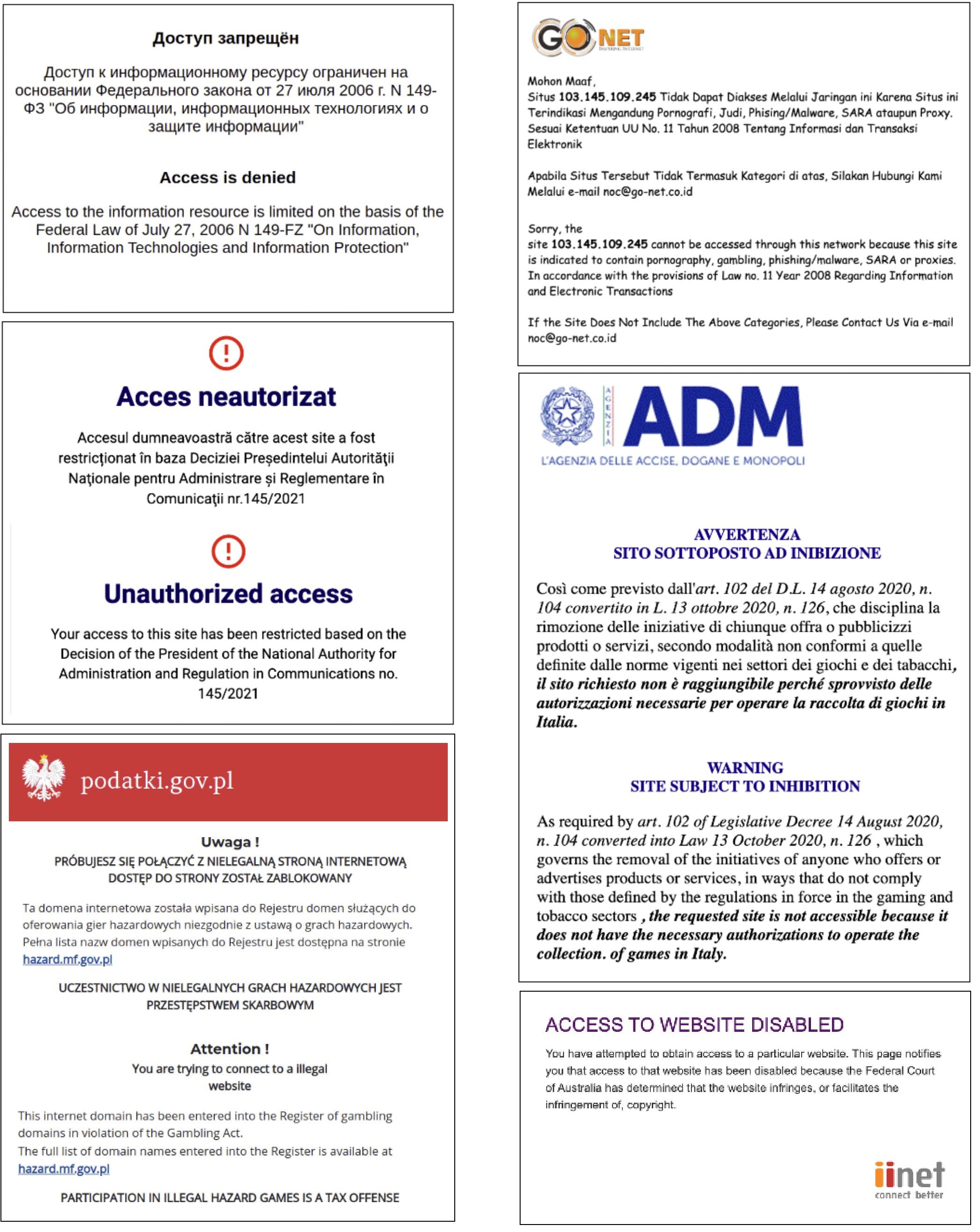}
    
  \caption{\textbf{ISP HTTP \bps detected by \platform} - Government \bps of Russia, Indonesia, Romania, Italy, Poland, and Australia}
\label{fig:http_only_isp_bp}
\end{figure*} 
%Figure~\ref{fig:http_only_isp_bp} shows examples of ISP blockpages. 

\end{document}